\documentclass{elsart}
\usepackage{amssymb}
\usepackage{graphicx}
\usepackage{epsfig}

\begin{document}
\begin{frontmatter}

\title{CUORE: A CRYOGENIC UNDERGROUND OBSERVATORY FOR RARE EVENTS}

\author[label1] {C. Arnaboldi},\author[label2] {F. T. Avignone III},
\author[label3,label4] {J. Beeman},\author[label5] {M. Barucci},
\author[label6] {M. Balata},\author[label1] {C. Brofferio},
\author[label6] {C. Bucci},\author[label7] {S. Cebrian},
\author[label2] {R. J. Creswick},\author[label1] {S. Capelli}, 
\author[label1] {L. Carbone},\author[label1] {O. Cremonesi}, 
\author[label8] {A. de Ward},\author[label1] {E. Fiorini}, 
\author[label2] {H. A. Farach},\author[label8] {G. Frossati}, 
\author[label9] {A. Giuliani},\author[label1] {D. Giugni}, 
\author[label1] {P. Gorla},\author[label3,label4] {E. E. Haller},
\author[label7] {I. G. Irastorza},\author[label3] {R. J. McDonald},
\author[label7] {A. Morales},\author[label3] {E. B. Norman},
\author[label1] {P. Negri}, \author[label1] {A. Nucciotti},
\author[label9] {M. Pedretti},\author[label7] {C. Pobes},
\author[label10] {V. Palmieri}, \author[label1] {M. Pavan},
\author[label1] {G. Pessina}, \author[label1] {S. Pirro},
\author[label1] {E. Previtali}, \author[label2] {C. Rosenfeld},
\author[label3] {A. R. Smith}, \author[label1] {M. Sisti},
\author[label5] {G. Ventura}, \author[label1] {M. Vanzini},
\author[label1] {L. Zanotti}

\address[label1]{Dipartimento di Fisica dell'Universit\`{a} di 
Milano-Bicocca e Sezione di Milano dell'INFN, Milan I-20136,
Italy} 

\address[label2]{University of South Carolina, Dept.of Physics
and Astronomy, Columbia, South Carolina, USA 29208 }

\address[label3]{Lawrence Berkeley National Laboratory, Berkeley,
California, 94720, USA}

\address[label4]{Dept. of Materials Science and Mineral
Engineering, University of California, Berkeley, California
94720, USA }

\address[label5]{Dipartimento di Fisica dell'Universit\`{a} di
Firenze e Sezione di Firenze dell'INFN, Firenze I-50125, Italy }

\address[label6]{Laboratori Nazionali del Gran Sasso, I-67010,
Assergi (L'Aquila), Italy}

\address[label7]{Laboratorio de Fisica Nuclear y Altas Energias,
Universid\`{a}d de Zaragoza, 50009 Zaragoza, Spain }

\address[label8]{Kamerling Onnes Laboratory, Leiden University,
2300 RAQ, Leiden, The Netherlands }

\address[label9]{Dipartimento di Scienze Chimiche, Fisiche e
Matematiche dell'Universit\`{a} dell'Insubria e Sezione di Milano
dell'INFN, Como I-22100, Italy }

\address[label10]{Laboratori Nazionali di Legnaro, Via Romea 4,
I-35020 Legnaro ( Padova ), Italy }

\centerline{(The CUORE COLLABORATION)}

\begin{abstract}
{CUORE is a proposed tightly packed array of 1000 $TeO_{2}$
bolometers, each being a cube 5 $cm$ on a side with a mass of 760
$g$. The array consists of 25 vertical towers, arranged in a
square of 5 towers by 5 towers, each containing 10 layers of 4
crystals. The design of the detector is optimized for ultralow-background 
searches: for neutrinoless double beta decay of $^{130}Te$ 
(33.8\% abundance), cold dark matter, solar axions,
and rare nuclear decays. A preliminary experiment involving 20
crystals $3\times3\times6\;cm$ of $340\;g$ has been completed, and a
single CUORE tower is being constructed as a smaller scale
experiment called CUORICINO. The expected performance and
sensitivity, based on Monte Carlo simulations and extrapolations
of present results, are reported.}
\end{abstract}

\end{frontmatter}

\section{Introduction} \label{sec:intro}

Neutrinoless double-beta decay, is a process by which two neutrons in a nucleus might 
beta decay by exchanging a virtual Majorana neutrino, and each  emitting an electron.
This would violate lepton number conservation $(\Delta L = 2)$ \cite{furry}, and would
require neutrinos to be Majorana particles. There are many reviews on the subject 
\cite{primakoff,haxton,morales,elliottvogel} 
The decay rate for the process involving the exchange of a Majorana neutrino can be 
expressed as follows:
\begin{equation}
	\frac{1}{\tau^{0\nu}_{1/2}}=
 	G^{0\nu}(E_0,Z)\;\frac{|< m_\nu >|^{2}}{m_e^2}\;|M^{0\nu}_f-( g_A /g_V )^{2}M^{0\nu}_{GT}|^{2}.
	\label{eq:lambda}
\end{equation}
In equation (\ref{eq:lambda}), $ G^{0\nu} $ is the two-body phase-space factor including coupling 
constants, $ M^{0\nu}_f $ and $ M^{0\nu}_{GT} $ are the Fermi and Gamow-Teller nuclear matrix 
elements respectively, and $g_A$ and $g_V$ are the axial-vector and vector relative weak coupling 
constants, respectively and $m_e$ is the electron mass. The quantity $|<m_\nu >|$ is the effective Majorana electron neutrino 
mass given by:
\begin{equation}
    |< m_\nu >|\equiv \Big| \;|U^{L}_{e1}|^{2}m_{1} + |U^{L}_{e2}|^{2}m_{2}e^{i\phi_{2}} + 
    |U^{L}_{e3}|^{2}m_{3}e^{i\phi_{3}} \; \Big|,
\end{equation}
where $ e^{i\phi_{2}} $ and $ e^{i\phi_{3}} $ are Majorana CP phases ($ \pm1 $ for CP conservation) 
and $ m_{1,2,3}$ are the mass eigenvalues.

The electron neutrino, for example, is a linear combination of neutrino mass-eigenstates i.e.: 
\begin{equation}
	|\;\nu_{e}>\; =\; U^{L}_{e1}\;|\;\nu_{1}> + \;U^{L}_{e2}\;|\;\nu_{2}> 
	+ \;U^{L}_{e3}\;|\;\nu_{3}>.
\end{equation}
The effective Majorana neutrino mass, $|< m_\nu >|$, is directly
derivable from the measured half-life of the decay as follows:
\begin{equation}
	|< m_\nu >| = m_e \; \frac{1}{\sqrt{ F_N \,\tau^{ 0\nu}_{ 1/2 }\,}}\; \; [eV],
\label{eq:mnu}
\end{equation}
where $ F_N \equiv G^{ 0\nu }\,| M^{ 0\nu}_f - ( g_A / g_V )^{2}\, M^{0\nu }_{ G T } |^{2}$. 
This quantity derives from nuclear structure calculations and 
is model dependent as shown later.

The most sensitive experiments thus far utilize germanium detectors isotopically enriched in 
$ ^{76}Ge $ from  7.8\%  abundance to $\sim$ 86\% . This activity began with natural
abundance $ Ge $ detectors by Fiorini {\it et al.} in Milan \cite{fiorini}
evolving over the years to the first experiments with small
isotopically enriched $Ge$ detectors \cite{vasenko}, and finally to
the two present multi-kilogram isotopically enriched
$^{76}Ge$ experiments: Heidelberg Moscow \cite{bandis} and IGEX \cite{aalseth,ceaalseth}.

Where should the field of $ \beta\beta $ - decay go from
here ? Suppose we consider the observed neutrino oscillations
in the data from atmospheric neutrinos \cite{fukuda} and solar
neutrinos \cite{hampel}. These, coupled with the recent exciting results
reported by the Sudbury Neutrino Observatory (SNO) Collaboration \cite{ahmad},
gives clear evidence that solar neutrinos oscillate,
and that the standard solar model is correct, but that solar electron neutrinos are
oscillating to other flavors on their way from the sun, or in the sun. In addition,
neutrino oscillation experiments have yielded the following values 
of $ |U_{ek}|^{2} : |U_{e1}|^{2} = (0.75^{+0.07}_{-0.12}) $, 
 $ |U_{e2}|^{2} = (0.25^{+0.12}_{-0.07}) $, and 
 $ |U_{e3}|^{2} = (<0.026) $ with confidence levels of 3 $\sigma$.

A number of recent theoretical interpretations of these data
\cite{pascoli,farzan,bilenky,mbilenky,klapdor,sbilenky} imply
that the effective Majorana mass of the electron neutrino, $ |< m_{\nu} >| $, expressed in
Eq. (2), could be in the range 0.01 $eV$ to the present bounds \cite{bandis,ceaalseth}. 
Considering this range, could
a next generation $ 0\nu\beta\beta $ - decay experiment detect it? If so,
what technique would be the best for a possible discovery
experiment? We will address these questions in an effort to demonstrate that CUORE,
an array of $1000$, $760\; g$ $TeO_2$ bolometers, is one of the best approaches
presently available. It can be launched without isotopic
enrichment nor extensive  Research and Development (R \& D) , and it can achieve next generation sensitivity.

The CUORE project, discussed in Sec.~\ref{sec:project} originates as a natural 
extension of the succesfull MI-DBD $^{130}Te$ experiment \cite{mibetafinale} where
for the first time a large array of bolometers was used to search 
for $ 0\nu\beta\beta $ - decay. The good results obtained so far prove that the \emph{bolometric}
technique although novel is competitive and alternative to the traditional
\emph{calorimetric $Ge$} technique. In Sec.~\ref{sec:project} details of the
detectors of the CUORE array are discussed while in Sec.~\ref{sec:cryogenic} the cryogenic system 
required to cool the array at the operating temperature of $\sim 10\;mK$ is described. 
Background issues are considered in Sec.~\ref{sec:radioac} while electronics, DAQ and data-analysis
are discussed in Sec.~\ref{sec:electronics} and Sec.~\ref{sec:analysis}.
The background simulation and the predicted sensitivity for CUORE are analysed in Sec.
\ref{sec:simulation} where we also discuss briefly the physical potential of CUORE regarding 
Dark Matter and solar axions searches. In Sec.~\ref{sec:enrich} the enrichment option is 
discussed.

\section{The CUORE project} \label{sec:project}

\begin{figure}
 \begin{center}
 \includegraphics[width=0.95\textwidth]{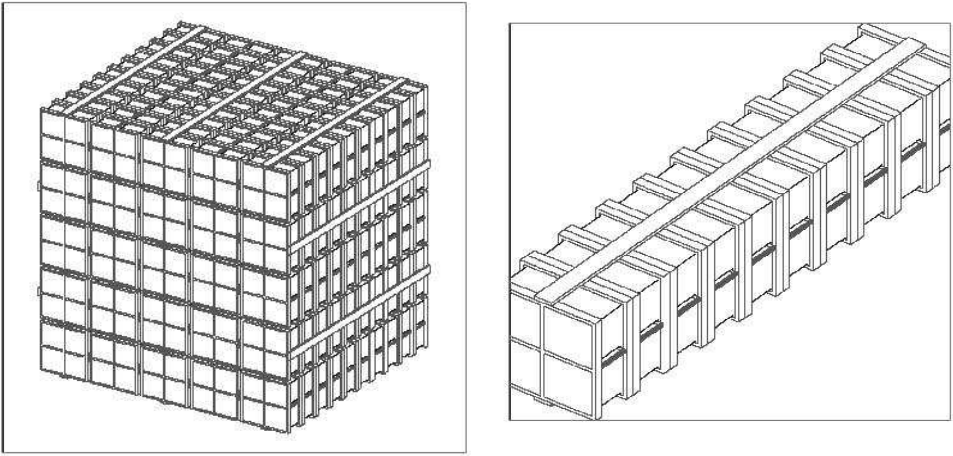}
 \end{center}
 \caption{The CUORE detector (left), one of the 25 towers (right).}
 \label{cuore_cubo}
\end{figure}

The CUORE detector will consist of an array of 1000 $TeO_{2}$
bolometers arranged in a square configuration of 25 towers of 40
crystals each (Fig.~\ref{cuore_cubo}). The principle of operation of these bolometers is
now well understood \cite{twerenbold}. Tellurium Oxide is a dielectric and
diamagnetic material. According to the Debye Law, the heat capacity of a single
crystal at low temperature is proportional to the ratio $ (T/T_{\Theta})^{3} $, 
where $T_{\Theta}$ is the
Debye Temperature of $TeO_{2}$. Thus, providing that the temperature is 
extremely low, a small energy release in the crystal results in to a measurable
temperature rise. This temperature change can be recorded with thermal sensors and 
in particular
using Neutron Transmutation Doped (NTD) germanium thermistors.
These devices were developed and produced at the Lawrence Berkeley 
National Laboratory (LBNL) and UC Berkeley Department of Material Science \cite{haller},
they have been made unique in their uniformity of response and 
sensitivity by neutron exposure control with neutron absorbing foils
accompanying the germanium in the reactor \cite{norman}.

The $ TeO_{2} $ crystals are produced by the Shanghai Quinhua
Material Company (SQM) in Shanghai, China and they will be the source of
760 $g$ $ TeO_{2} $ crystals for CUORE \cite{dubna}.
A single CUORE detector consists of a $5\times5\times5\;cm^{3} $
single crystal of $TeO_{2}$ that acts both as a detector and source. 
The detectors will be supported in a copper frame in the 25 tower 
configuration shown in Fig.~\ref{cuore_cubo}.
The frame, and dilution refrigerator mixing chamber to which it
is thermally connected, forms the heat sink, while the PTFE (Polytetrafluoroethylene or TEFLON)
stand-offs provide the thermal impedance which delays the
re-cooling of the bolometers. The bolometers operate at 
$\sim 10\;mK$.

A single tower of CUORE is presently under construction. It will be attached 
to the mixing chamber of the same dilution refrigerator (DR) used in the Milano 
20 crystal array experiment (MI-DBD) and run not only as a test bed for 
CUORE but will also be an experiment called CUORICINO \cite{dubna,cuoricino} (which in 
Italian means small CUORE) designed to improve 
on the present sensitivity of $ |< m_\nu >| $ obtained with isotopically enriched 
$Ge$ detectors \cite{bandis,aalseth}. CUORICINO will prove the feasibility of the 
extension of the MI-DBD technology to large arrays. 
The smaller array of 20,  $3\times
3\times 6 \;cm^{3}$ crystals was operated successfully for
31,000 $hours \times kg$. The data yield a bound $ \tau^{0\nu}_{1/2}(^{130}Te)
> 2 \times 10^{23}\;y $, corresponding to $ \mid < m_{\nu} > \mid < (1\; or\; 2)\;eV $, 
depending on the nuclear matrix element used \cite{mibetafinale}.

The CUORE detector then will be
the final version of a three step development comprising: MI-DBD,
CUORICINO, and finally CUORE. This, plus the fact that CUORE
requires no isotopic enrichment, (the isotopic abundance of 
the  $\beta \beta $ emitter  $ ^{130}Te $ is 33.8\%) puts CUORE ahead 
of other options of truly next generation $ 0 \nu \beta \beta $
experiments. The technology, though novel, is developed and
to a large degree proven.

\subsection{The single CUORE detector} \label{sec:single}

The single CUORE detector is a a $5 \times 5 \times 5 \;cm^{3} $
single crystal of $ TeO_{2} $ grown with ultrapure $ TeO_{2} $ 
powders and optically polished on the surfaces.
Crystals of $ TeO_{2} $ have a tetragonal $ <110> $ structure, and
are grown along the (001) axis. The two axes normal to this axis
are crystallographically equivalent, a fact relevant to their use
in the search for solar axions discussed later \cite{creswick}. The surface
hardness is not the same for all sides, which complicates the crystal
polishing. We have shown that repeated thermal cycling does not
damage the crystals, as in the cases of crystals of other
tellurium compounds, or those of tellurium metal.
The Debye temperature of the $ TeO_{2} $ crystals was specially
measured for the CUORE project as 232 $K$ \cite{bariucci}. This differs from the
previous value of 272 $K$ in the literature \cite{white}. The specific heat
of the $ TeO_{2} $ crystals follows the Debye law down to 60 $mK$;
the heat capacity of 760 $g$ crystals extrapolated down to 10 $mK$ is $ 2.3\times 10^{-9}\;
J/K $.

The NTD thermistors are attached by epoxy to the crystal and are operated in 
the Variable Range Hopping (VRH) conduction regime with a Coulomb gap
\cite{giuliani,mott,shklovskii}. 
The most important parameter characterizing the thermal response is the sensitivity, 
$ A $, defined as follows:
\begin{equation}
	A = \left|\frac{d \log R}{d \log T}\right| = \gamma \left( \frac{To}{T}\right)^{\gamma}.
	\label{eq:A}
\end{equation}
For our NTD thermistors this parameter ranges between 7 and 10. The resistance behavior follows the relation,
\begin{equation}
 	R = R_{0} \exp (T_{0}/T)^{\gamma}; \hskip1cm \gamma = 1/2.
\end{equation}

The VRH regime occurs in $Ge$ when it is ''doped'' close to the Metal to Insulator 
Transition, which is $ \sim 6\times 10^{16}\;atoms / cm^{3} $. This is achieved by 
thermal neutron radiation in a nuclear reactor, resulting in $ (n,\gamma) $
reactions on $ ^{70,72,73,74,76}Ge $, followed by $ \beta^{-}$-decay or electron capture. 
To first order, this produces (for a natural $Ge$ sample) the following stable isotopes: 
$ ^{71}Ga$(18\%), $ ^{73}Ge$(8.3\%), $^{74}Ge$(36\%), $^{75}As$(12\%), and $ ^{77}Se$(0.8\%). 
Beta minus and electron capture lead to n and p dopants, respectively.
The sensitivity parameter, $A$, depends on the neutron irradiation dose. Therefore, each 
thermistor must be characterized at operating temperatures, as described for $Si$ thermistors by
Alessandrello {\it et al.} \cite{aalessandrello}.

It is very important to optimize the neutron irradiation exposure and to make the exposures 
as uniform as possible. It is not possible to evaluate thermistor material directly from 
the reactor because of the long half life of $ ^{71}Ge $ (11.43 days). A delay, of several months, 
is required to see if the $Ge$ needs more exposure. To circumvent this difficulty, the $Ge$
material is accompanied by foils of metal with long-lived $(n,\gamma)$ radioactive daughter nuclides. Accordingly, the
neutron exposure of the $Ge$ can be determined accurately, and uniformity of exposure is achieved. 
This technique was developed recently by the Lawrence Berkeley National Laboratory group of the
CUORE Collaboration \cite{norman}. Following the neutron exposure and radioactive 
decay period, the NTD germanium is first heat treated to repair the crystal structure
then cut into $ 3\times 3\times 1 \;mm^{3} $ strips. 
The thermistors are glued to the $ TeO_{2} $ crystal by 9 spots of Araldit rapid, Ciba
Geigy epoxy, of 0.4 to 0.7 $mm$ deposited on the crystal surface by an array of pins.
The height of each spot is $ 50 \;\mu m $. This procedure was found
to be reliable and reproducible in the MI-DBD experiment \cite{mibetafinale}.
The heat conductance of the epoxy spots was measured in Milan and
the phenomenological relation was found to be $ \sim 2.6 \times
10^{-4} \;(T[K])^{3}$ watts per degree kelvin per spot.

The stabilization of the response of bolometers is crucial because of the unavoidable small 
variations in the temperature of the heat bath that change the detector gain (and consequently 
deteriorates the energy resolution). 
This problem is successfully addressed by means of a Joule heater glued on to each crystal. 
The heater is used to inject a uniform energy in the crystal, the thermal gain is monitored 
and corrected off-line (see Sec.~\ref{sec:analysis}). 
The heaters are $ Si $ chips with a heavily doped meander structure with a constant 
resistance between $ 50 $ to $ 100 \;k\Omega $. They are manufactured by the ITC - IRST 
company in Trento, Italy.

Electrical connections are made with two $ 50 \;\mu m $ diameter gold wires, ball bonded to 
metalized surfaces on the thermistor. The gold wires are crimped into a copper tube, which is 
inserted into a larger one forming the electrical connection, and avoiding low temperature solder 
which contains $ ^{210}Pb $ and traces of radioisotopes. The larger copper tube, $ \sim 14 \;mm $ 
long and $ 2 \;mm $ in diameter, is glued to the copper frame that supports the crystals. 
This tube is thermally connected to the frame but electrically insulated.

The mounting of the $ TeO_{2} $ crystals is crucial to detector
performance, and must fulfil a number of sometimes contradictory
criteria:
\begin{enumerate}
\item{crystals must be rigidly secured to the frame to prevent
power dissipation by friction caused by unavoidable vibrations,
that can prevent the crystal from reaching the required
temperature and can produce low frequency noise;}

\item{the thermal conductance to the heat sink (copper frame) must be
low enough to delay the re-cooling of the crystal, following a
heat pulse, such that the pulse decay time (re-cooling time) is
much longer than the rise time;}

\item{however, the heat conductance must be high enough to guarantee
efficient cooling;}

\item{the frame must compensate for the differences in thermal
expansion coefficients of the various materials used;}

\item{and finally, only materials selected for low radioactivity can
be used.}
\end{enumerate}
For CUORE, only two materials will be used, copper and PTFE;
they can both be obtained with very low levels of radioactivity.
Copper has a thermal conductivity and specific heat high enough to be an
ideal heat bath, and has excellent mechanical properties; it
machines well, and has excellent tensil, tortional and
compressional strength.

PTFE is used between the copper frame and the crystals. It has
low heat conductance and low heat leak \cite{pobell}. It compensates for the
differences between coefficients of thermal expansion of copper
and of $ TeO_{2} $. The heat conductance between the crystal and
heat sink was measured in a similar configuration, but with a
contact area larger by a factor of 2 \cite{rello}.

The measurements of the temperature dependence of specific heats
and thermal conductivities is crucially important for developing a
model that explains detector performance and that can be used to improve the
detector design.

\subsection{The detector model}\label{sec:model}

A rough approximation of the NTD thermistor voltage pulse amplitude $\Delta V_{bol}$ produced by an 
energy deposition $E$ into the $TeO_{2}$ absorber can be obtained by:
\begin{equation}
 \Delta V_{bol} = V_{bol} \cdot A \cdot \frac{\Delta T_{bol}}{T_{bol}} 
   = V_{bol} \cdot A \cdot \frac{E}{C\;T_{bol}}
 \label{eq:pulse}
\end{equation}
where $V_{bol}$ is the bias of the NTD thermistor, $ A $ is the sensitivity obtained from equation \ref{eq:A}, 
$\Delta T_{bol}$ is the temperature increase and $C$ is the heat capacity of the absorber. 
Using the heat capacity discussed earlier, a one $MeV$ 
energy deposition into the $TeO_{2}$ absorber results in $\Delta T_{bol} = 7\times 10^{-5}K$. Using
equation \ref{eq:pulse}, this results in $\Delta V_{bol} \sim 600 \; \mu V$; however, the
observed pulses are lower by a factor $\sim 5$, therefore a more quantitative
model is needed.

\begin{figure}
 \begin{center}
 \includegraphics[width=0.6\textwidth]{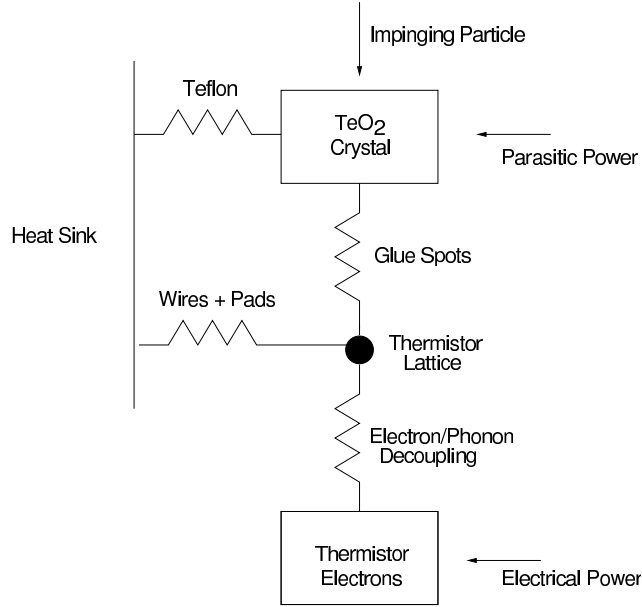}
 \end{center}
 \caption{Network representing the detector model.}
 \label{network}
\end{figure}

A semi-quantitative Thermal Model (TM) has been developed which describes the detector
in terms of three thermal nodes with finite heat capacities of the $TeO_{2}$ crystal absorber,
$Ge$  thermistor lattice,and $Ge$ thermistor electrons. For simplification, the lattice component
of heat capacity is assumed to be negligible. The network describing the detector is shown in
Fig.~\ref{network}. The two heat capacities, and 4 thermal conductances, the parasitic power
from the vibrations and power dissipated in the thermistor electrons, are the model parameters.
A typical energy deposition of 1~$MeV$ in the crystal is used and it is assumed that it is
instantaneously thermalized.

A computer code was developed in Milan to determine the static behavior. The required
input data are: the heat sink temperature, the temperature behavior of the thermal conductances
shown in Fig.~\ref{network}, the parasitic power in the crystal, and a set of values of electrical
joule power dissipated in the thermistor electrons. The code determines a complete solution
to the static problem, for each power level, and yields the temperature of the three thermal nodes,
the resistance of the thermistor, and the current through it. These parameters  define
the ``operational point''. A set of operational points, with a fixed heat sink temperature,
defines a ``load curve''. Load curves are expressed as a set of voltage-current, or power resistance points.
When the static operation point is determined, the time evolution of the pulse is
computed. 
The main aim of the TM is to have a reliable prediction of the detector figure 
of merit on the basis of the known thermal parameters. To address this question it is necessary to 
solve the non trivial problem of how it is possible to compare the performances of two detectors
The more sensitive thermistors yield larger amplitude pulses, but they are also more sensitive to spurious
noise, for example microphonics, crosstalk, and slow heat pulses from vibrations that
excite the 1-5 $Hz$ portion of the noise spectrum. More gain at the heat sink temperature
yields larger pulses at the price of higher resistance and more spurious noise.

Recent experience shows that the best way to define a detector figure of
merit is by fixing the heat sink temperature between $ 7\; mK $ and $ 11\; mK $ and by measuring
a load curve. Using particle or heater pulses, the operation point on the load curve,
corresponding to the highest signal, is determined. For each heat sink temperature,
one determines a pair of points, thermistor resistance, $R$, and signal amplitude, $V$, usually
expressed in microvolts per $MeV$, corresponding to the maximum pulse amplitude for a
given heat sink temperature. One obtains a characteristic curve for that detector by
plotting voltage vs resistance, which to a good approximation is a straight line on a
log-log plot. This is called the Detector Merit Curve (DMC). Experience shows that a
detector has very similar signal-to-noise ratio along a DMC.  No large differences in
performance are noted for heat sink temperatures of $8\; mK$ or $12\; mK$, because signal level
and the spurious noised offset one another. It is better to avoid the extremities of
the DMC that lead to higher spurious noise or low signals.

The best way to compare two different detectors is by using their DMCs. If the DMC
of detector A lies systematically above that of detector B, the performance of detector
A is superior. It provides the highest signals for a fixed thermistor resistance value.
The main value of the TM is to predict the DMC given the temperature behavior
of the thermal parameters discussed above.

\begin{figure}
 \begin{center}
 \includegraphics[width=0.95\textwidth]{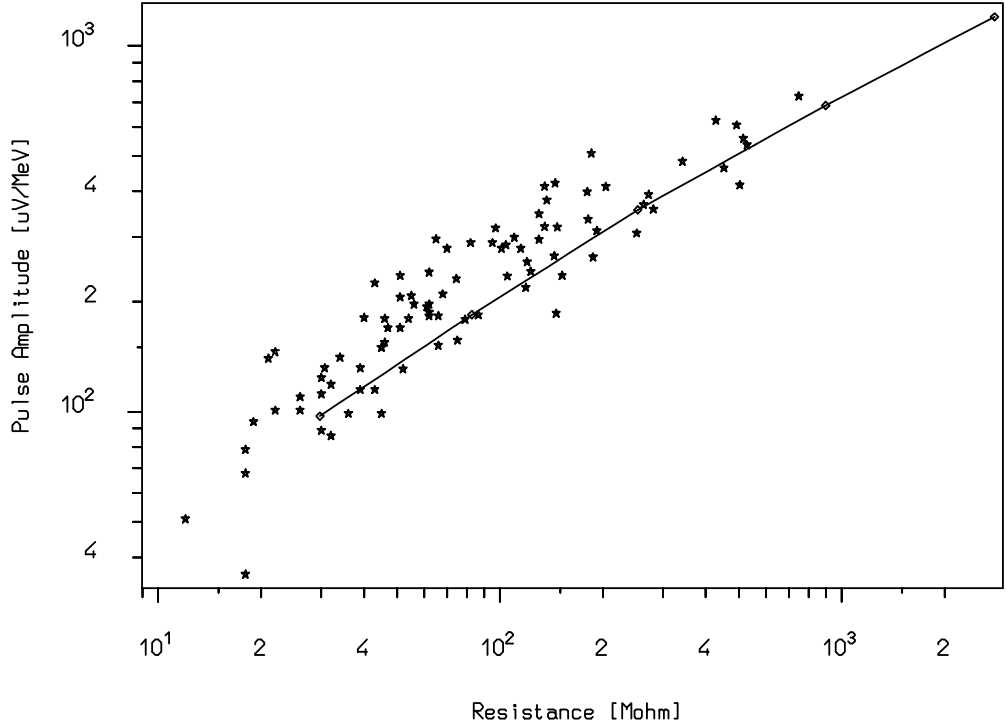}
 \end{center}
 \caption{Detector Merit Curve: simulation and experimental
 points for the 20 detector array.}
 \label{DMC}
\end{figure}

As a test of the model, a typical detector of the MI-DBD 20-element array was
simulated. These have different masses and geometries than the CUORE crystals. A DMC
was then constructed using the model for this detector. It was then compared to
the real DMC measured for the 20-detector array. A significant spread among the
experimental points, for all 20-detectors was found. There were two reasons for
this ,intrinsic detector differences and the presence of points not corresponding
exactly to the maximum signal. Nevertheless the simulated DMC lies inside of the
experimental points and has the correct slope (see Fig.~\ref{DMC}). The detector
simulations are not distinguishable from a typical detector array.

\begin{figure}
 \begin{center}
 \includegraphics[width=0.95\textwidth]{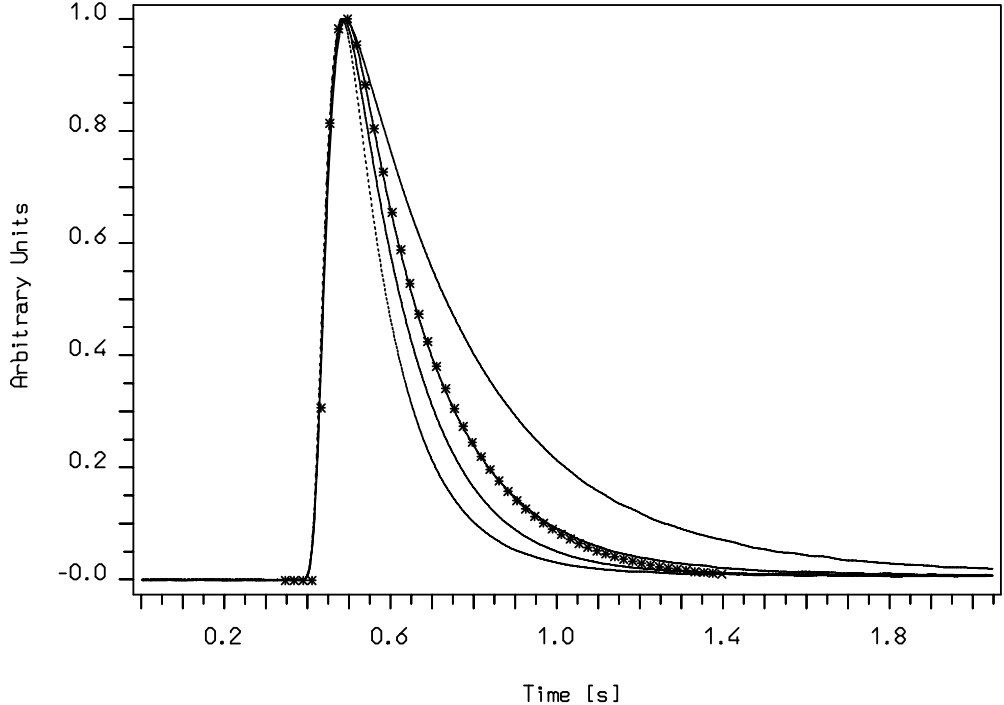}
 \end{center}
 \caption{Real pulses (full line) and simulated pulses (line+points).}
 \label{pulse}
\end{figure}

The TM must correctly predict not only pulse amplitude, but also time
constants. At a typical operating resistance ($\sim 100\, M \Omega$) the predicted rise time
(10\%-90\%) and decay time, are respectively 40~$ms$ and 430~$ms$, and lie inside the
distribution of these parameters for the 20-detector array. In Fig.~\ref{pulse}, the
simulated pulse is compared with real pulses collected with four detectors, randomly
chosen among the 20 elements. It is evident that the pulse-shape of the simulated
detector is very similar to that of a typical array of detectors. From this it was
concluded that the TM is capable of explaining semi-quantitative
performance of the 20 - element array.

The TM can be used in parallel with specific low temperature experimental
tests, to define the optimum properties of the CUORE element, and the results thus
obtained imply the following  optimum parameters for the CUORE elements:
\begin{enumerate}
\item{$TeO_2$ absorber dimensions: $50 \times 50 \times 50\;mm$, of mass of 760~$g$;}

\item{NTD Thermistor \# 31, $T_0=3.0\,K$, $R_0=1.5\,\Omega$, dimensions $3\times 3\times 1$~$mm$,
     with a contact distance of $3\,mm$, and with a resistance of $50\,M\Omega$ 
     at $T=10\,mK$; }
\item{absorber-thermistor coupling made with 9 epoxy spots, 0.5 to 0.8 $mm$ in 
diameter, and 50 $\mu m$ high;}
\item{gold connecting wires 50 $\mu m$ in diameter, and 15 $mm$ in length;}
\item{crystals mounted on PTFE blocks with a total
    PTFE-to-crystal contact area equal to about a few square $mm$.}
\end{enumerate}

\subsection{The modular structure of the CUORE detector}\label{sec:modular}

The CUORE array comprises 1000 $ TeO_{2} $ bolometers, grouped in 250 modules of 4 bolometers each. 
These are arranged in 25 towers of 40 crystals each, or a stack 10 modules high. The towers are
assembled in a 5 by 5 matrix as shown in Fig.~\ref{cuore_cubo}.

\begin{figure}[bt]
    \begin{minipage}[c]{1\textwidth}%
      \centering
      \includegraphics[width=0.5\textwidth]{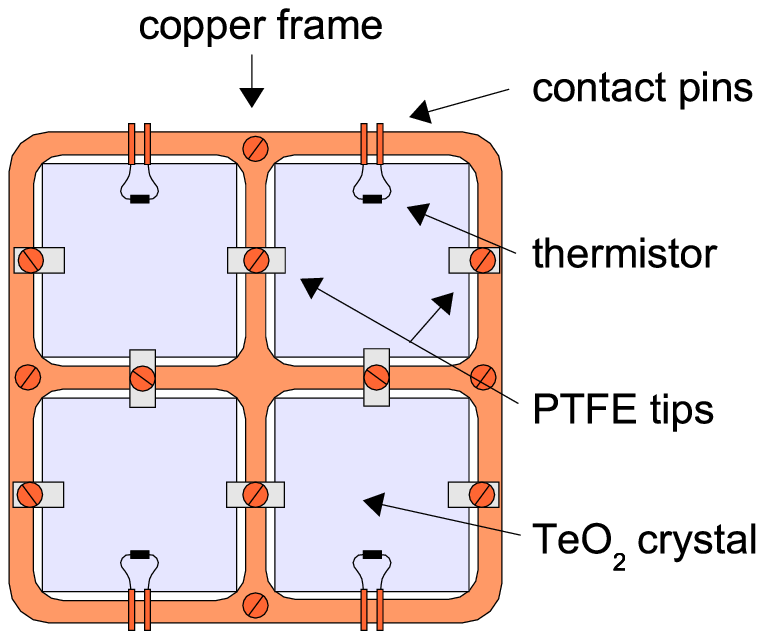}%
      \includegraphics[width=0.5\textwidth]{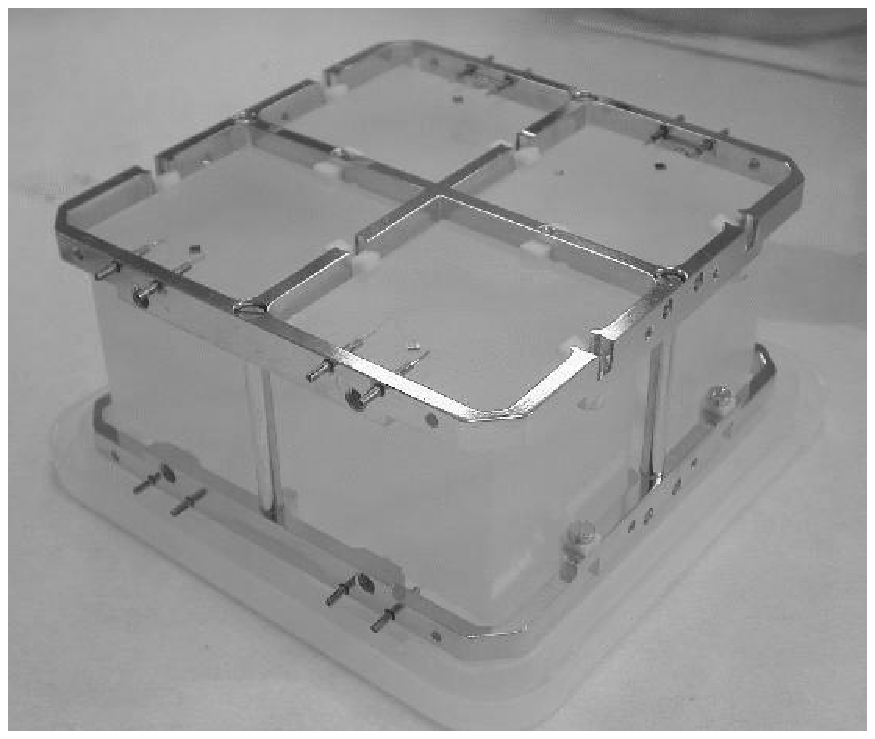}%
    \end{minipage}%
 \caption{A four detector module.}
\label{piano}
\end{figure}

The structure of each four-detector module is shown in Fig.~\ref{piano}. The four crystals are held 
between two copper frames joined by copper columns. PTFE pieces are inserted between the copper 
and $ TeO_{2} $, as a heat impedance and to clamp the crystals. 
There is a 6 $mm$ gap between crystals with no material between them. 

\begin{figure}
 \begin{center}
 \includegraphics[width=0.95\textwidth]{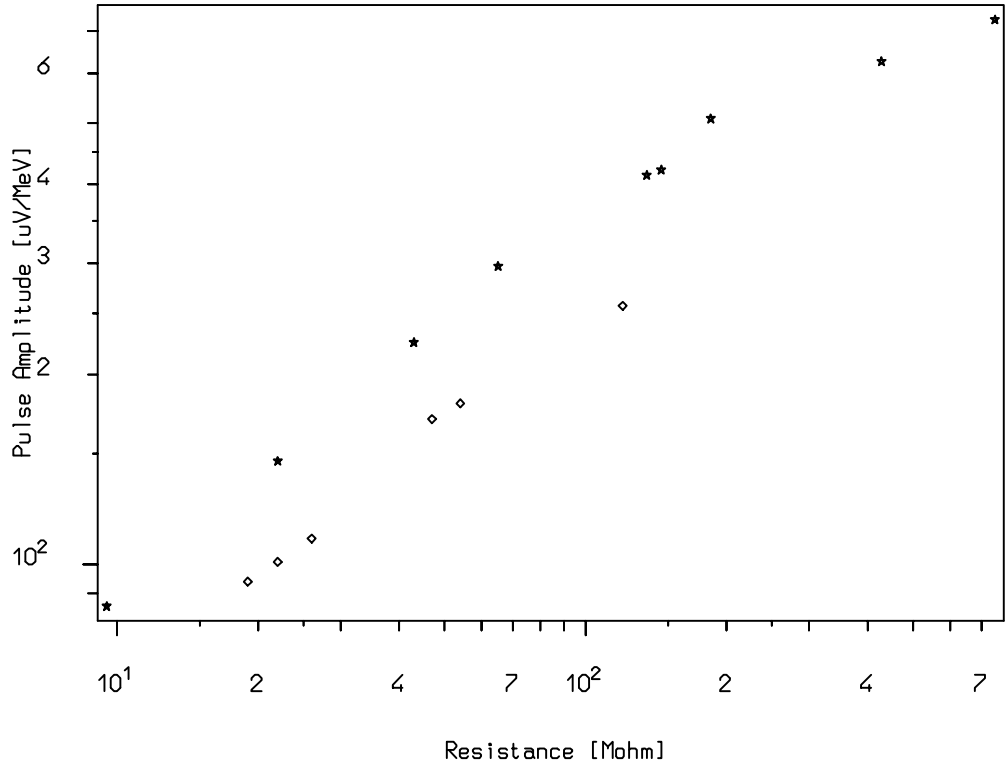}
 \end{center}
 \caption{Improvement in the Detector Merit Curve by increasing the thermistor
 size: $ 3 \times 3 \times 1 \;mm^{3} $ thermistors (stars) vs. 
 $ 3\times 1.5 \times 0.4 \;mm^{3} $ (diamonds).}
 \label{DMC-2}
\end{figure}

\begin{figure}
 \begin{center}
 \includegraphics[width=0.95\textwidth]{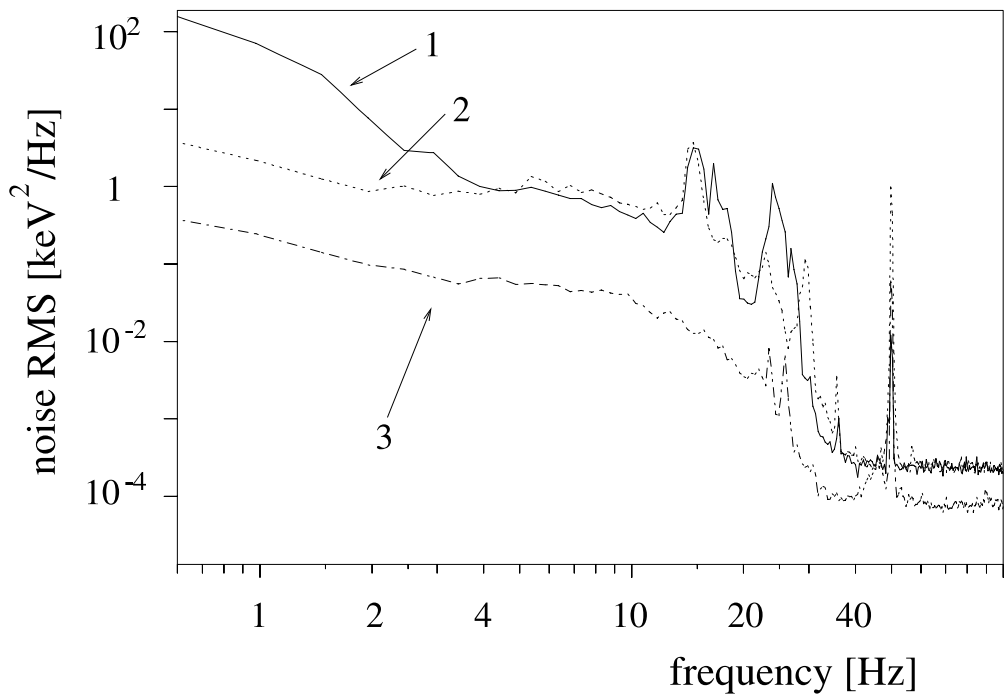}
 \end{center}
 \caption{Noise Power Spectra without damping suspension (1), with 
 damping suspension only (2) and with dumping suspension + cold electronics (3).}
 \label{noise}
\end{figure}

The four detectors are mechanically coupled; some of the PTFE
blocks and springs act simultaneously on two crystals. Tests on
this and a similar structure but with $3 \times 3 \times 6 \;mm^{3}$
crystals clearly demonstrate that the CUORE technology is
viable. Noise reduction was achieved by replacing the $ 3\times
1.5 \times 0.4 \;mm^{3} \; Ge $ thermistor by one with dimensions $ 3
\times 3 \times 1 \;mm^{3} $, which significantly improved the DMC shown in 
Fig.~\ref{DMC-2}, and by utilizing cold electronics Fig.~\ref{noise}. 
~

\begin{figure}
 \begin{center}
 \includegraphics[width=0.95\textwidth]{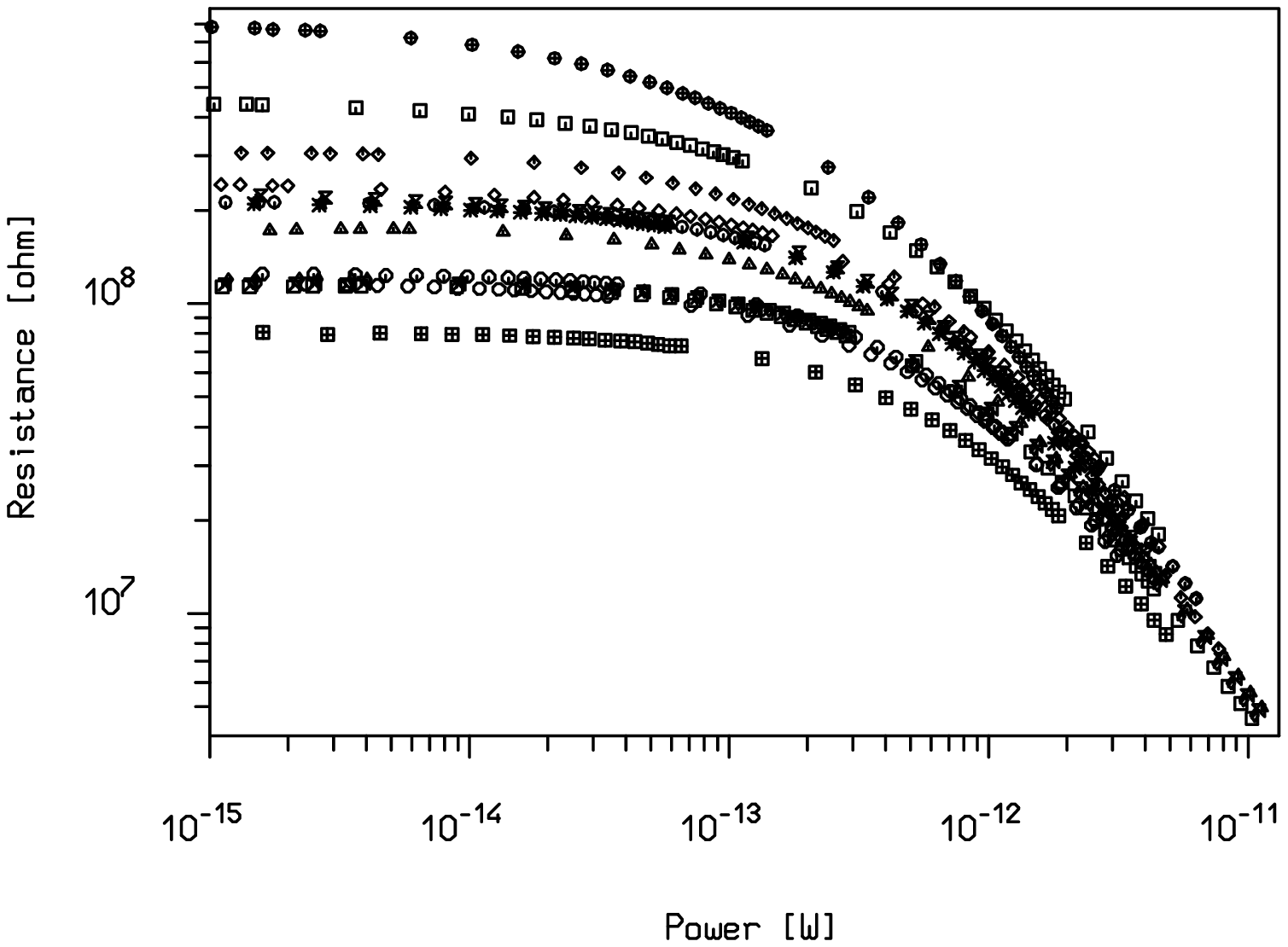}
 \end{center}
 \caption{Fifteen thermistor load curves for the 20 crystal array.}
 \label{load_curve}
\end{figure}

Reproducibility was tested on the MI-DBD array, which is a
significant number of detectors (20) operating simultaneously. The
reproducibility is satisfactory as shown in the load curves in
Fig.~\ref{load_curve}. Four-detector modules of $ 50 \times 50 \times 50 \,
mm^{3} $ crystals were also successfully tested, with pulse amplitudes spanning an
interval from 50 to 150 $ \mu V / MeV $ at $\sim 100\,M \Omega $ operation point, 
in agreement with the TM. The full width at half maximum
FWHM is $ \sim \,1\,keV $ for low energy gamma peaks, and $ \sim
\,5 \;to\; 10 \,keV $ at $2.6 \, MeV$.
A stack of 10, 4-detector modules, will be connected by two
vertical copper bars. The details of the wiring of the towers is
still under development; however, the current plan is for 50 $\mu
m$ diameter twisted pairs of nylon coated constantane wire,
running along the vertical copper bars. On the top of each tower, 
there would be 40 twisted pairs for the thermistors, and 4
for the heaters. Two more twisted pairs will be required for
diagnostic thermometers, one on top, and one on the bottom of the
tower. 

The CUORICINO detector presently under construction consists of
one of the 40 detector towers. CUORE is an array of 25 towers,
each of which will be suspended independently from a large square
copper plate, thermally connected to the mixing chamber of the DR.
In this manner, tests of CUORICINO will be tests of the CUORE
array. The array will be suspended from the mixing chamber by a vertical
spring to decouple the detector from the
vibrations of the dilution refrigerator. Tests were performed with
piezo-electric accelerometers and vibrators to optimize the
suspension (see Fig.~\ref{noise}). As a result a pendulum mechanical
suspension was designed for the entire CUORE array.

The heavy shielding of CUORE will present non-trivial problems in
the energy calibration with radioactive sources. The presently
considered option is with radioactive metal wires encapsulated in
PTFE tubes. The tubes would be placed vertically between the towers, 
and free to slide along a fixed tube
to the appropriate calibration point. A vacuum tight sliding seal
will allow the part of the tube containing wire to be inserted and
extracted from the cryostat.

\section{The cryogenic system} \label{sec:cryogenic}

The CUORE bolometers will operate at temperatures between 7 and 10
$mK$. This will require an extremely powerful dilution refrigerator (DR).
At these temperatures, the cooling power of a DR varies approximately as $
T^{2} $. Estimates were made of the parasitic power the detector
and DR would receive from: heat transfer of the residual helium
gas in the inner vacuum chamber (IVC), power radiated from the 50
$mK$ shield facing the detector, and from vibrational energy
(microphonic noise).

The estimated value is $ \sim 1 \; \mu W $ at 7 $mK$, using
reasonable values for the residual gas pressure and achievable
surface quality for radiation transfer. The resulting estimate for
the radiation contribution was negligible. CUORE will utilize a
similar system to that of the Nautilus Collaboration that cools a
2-ton gravitational antenna. That system experienced a parasitic
power of $ 10 \; \mu W $ from unknown sources.

\begin{figure}
 \begin{center}
 \includegraphics[width=0.5\textwidth]{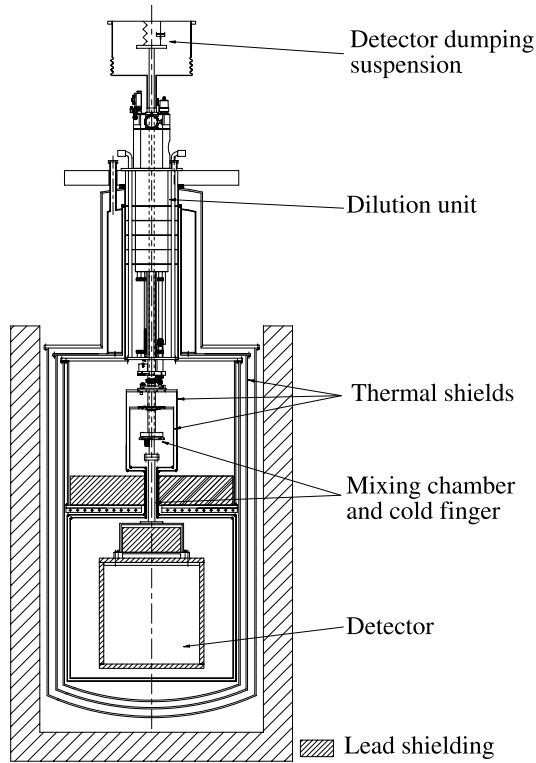}
 \end{center}
 \caption{CUORE cryostat and shielding.}
 \label{dewar}
\end{figure}

The CUORE detector will be cooled by a $He^3/He^4$ refrigerator with a cooling power of
$ 3 \;mW $ at $ 120 \;mK $. Refrigerators with the required characteristics are 
technically feasable, one example is the DRS-700 DR model
constructed by the Kamerling Omes Laboratory in Leiden. 
The unit is shown in Fig.~\ref{dewar}, inserted in the dewar. 
One important design
feature is the $ 50 \;mm $ diameter clear access to the mixing chamber
to allow a rod, suspended from an external structure, to suspend
the detector array to minimize vibrations from direct connection
to the mixing chamber. The temperature of the rod will be
controlled in stages along its length by flexible thermal contacts
to avoid vibrations.

The dewar housing the DR will have a jacket of liquid nitrogen (LN), to avoid the
need of superinsulation, which is not free of radioactivity. The
system is designed with several tubes passing through the LN
bath, the liquid $He$ bath, and the IVC, to allow refilling with
cryogenic liquids, and for sufficient feed through for electrical
connections. Liquifiers will provide constant liquid levels for
the two baths over long periods of operation.

The $He$ that evaporates from the main bath, and from the 1K-pot,
will be recovered in a buffer tank and re-liquified with a helium
liquifier with a capacity of several tens of $ l / day $. The gas will be
cooled to $ \sim 70\, K $, then to 10~$K$, by a two stage
Gifford-MacMahon cycle. This will be followed by a Joule-Thompson
cycle that liquifies it, and injects it back into the main bath.
The nitrogen liquifier, on the other hand, will require only a
single stage Gifford-MacMahon cycle to cool a cold finger above
the LN bath. It will condense the evaporated $ N_{2} $ and let it
fall back into the bath.

This complex cryogenic system will require a constant monitoring
system. It must be capable of surveying the operating conditions
to allow prompt intervention to correct malfunctions rapidly. In
addition, the operating conditions of the entire cryogenic system
must be recorded for future correlation with instabilities in the
experimental data.

\section{Radioactivity and shielding} \label{sec:radioac}

Double-beta decay experiments, as well as dark matter searches and
searches for rare events in general, require deep underground locations  
and ultra low radioactive environments. 
This latter requirement is usually accomplished by a severe selection of the
materials used and by the realization of proper shielding surrounding the 
detectors. In the case that complicated structures like a dilution refrigerator 
are used, the problem of shielding and of using only low radiactive 
contamination materials is complex.

\subsection{Shielding}

In the case of CUORE, part of the bulk lead shielding will be
placed inside of the cryostat, and part outside (Fig.~\ref{dewar}). 
This will accomplish shielding against the dewar, and will reduce the total
amount of lead required. A $ 4\pi $ layer of ultra-low background
lead will constitute 3 $cm$ thick walls of the cubic structure
of the array. This layer will be Roman lead whose $^{210}Pb$ activity was
measured to be less than  $4\; mBq / kg$ \cite{drllo}.
The dilution refrigerator will be constructed from materials
specially selected for low levels of radioactivity. Nevertheless,
these levels might be higher than can be tolerated by the
sensitivity requirements.

The top of the detector array will be protected by two layers 1 $m$ by 1 $m$, 
with a 10 $cm$ diameter central bore to accommodate 
the copper cold finger that supports the detector and the narrow neck of two
radiation shields of the refrigerator that are at temperatures of
50 $mK$ and 600 $mK$. The layer close to the detector will be 10 $cm$
thick, made from high quality lead with an activity of $16\; Bq /
kg$ of $ ^{210}Pb $. The upper layer, also 10 $cm$ thick, will be
made of modern lead with an activity of $ ^{210}Pb $ of $150\; Bq
/ kg$.  Another layer of lead 17 $cm$ thick, and 40 $cm$ by 40 $cm$ will be
placed directely on the top face of the detector. 
It will be constructed from low activity lead of $16\; Bq / kg$ of $ ^{210}Pb $. 
This configuration is designed so that the minimum path to the detector from the 
IVC and the dilution unit is 20 $cm$ of lead. Finally, outside the dewar, there will be 
two 10 $cm$ thicknesses of lead, $16\, Bq / kg$ of $ ^{210}Pb $ for the inner layer, and
$150\, Bq / kg$ for the outer layer.
The lead shield will be surrounded with a 10 $cm$ thick box of
borated polyethylene that will also function as an hermetically
sealed enclosure to exclude radon. It will be flushed constantly
with dry nitrogen. The 

The entire dewar, detector, and shield will be enclosed in a
Faraday cage to exclude electromagnetic disturbances that also
constitute a source of background, albeit at low energies,
important to dark matter and solar axion searches.
The addition of a muon veto surrounding the entire structure will be also considered.

\subsection{Material selection and treatment}

All the materials used to built the detectors, their mounting structure,
the cryostat and the shielding themselves will be selected to ensure 
that only low radioactive contamination materials will be empolyed.
A particular care will be obviously devoted to the detectors.
The $TeO_{2}$ crystals will be grown from ultrapure $TeO_{2}$ powders
to guarantee an extremely low level of bulk contamination (present limit 
on U and Th contamination is of the order of $\sim 10^{-13} g/g$).
A particlar care will be devoted to their surface treatment. 
Indeed the first crystals grown by SQM were polished in China with
conventional cerium oxide polishing rouge. This resulted in a
$ ^{238}U $ and $ ^{232}Th $ surface contamination of the crystals that
contribute to the measured background in the $0\nu\beta\beta$ of the Mi-DBD
first run. A re-polishing of the crystals in Italy with low contamination powders
reduced that contamination in the Mi-DBD second run.
A new polishing process studied in collaboration with SQM and a semi-authomatic 
polishing system realised in LBNL was used for the surface treatment of the 
CUORICINO crystals (partially done in China and completed in Italy). 
Also copper and PTFE used to construct the CUORE array will be selected for their
low contamination. Once machined all the copper and PTFE pieces will undergo
a surface cleaning procedure that will guarantee the required low level of surface
radioactive contamination for those parts that directely face the detectors.
The array will be assembled in a low $Rn$ clean room to avoid contamination from $Rn$ daughters.

\subsection{Location}

CUORE will be located in the underground halls 
of Laboratori Nazionali del Gran Sasso (L'Aquila - Italy) at a depth of 
3400 $m.w.e.$ where the muon flux is reduced to $\sim 3 \; 10^{-8} \mu/cm^{-2}/s$ 
and the neutron flux to $\sim 10^{-6} n/cm^{-2}/s$.

\section{Electronics and Data Acquisition} \label{sec:electronics}
\def \gohm{$ G \Omega$}
\def \ohm{$\Omega$}

The \gohm\ front-end electronics furnishes a bias current to the NTD
thermistors and receives and processes the resulting  signal-bearing voltage.
At the moment we are investigating two solutions that differ by the addition of 
a cold buffer stage inserted between the detector and the room temperature preamplifier 
in one case.
The connection from the thermistors to the electronics is via shielded twisted pair from 
one to five meters in length, depending by the final solution adopted.
The design of the electronics addresses four major issues.
First, it should minimize the noise generated by the biasing circuitry
and the preamplifier in the frequency band from $1\sim Hz$ to a few tens of $Hz$.
Second, it should accommodate the broad spread of manufacturing
characteristics that is typical of the bolometer/thermistor modules.
For some modules, for example, the optimum bias current is as low as
$150\;pA$ and for others as high as $300\;pA$.
As a consequence the signal amplitude exhibits a corresponding spread.
Third, it  should enable in situ measurement of the bolometer
characteristics.
Fourth, the design should incorporate a high level of remote
programmability in order that adjustments required during the observing
period do not entail mechanical interventions in the vicinity of the
bolometers.
In the following subsections we give a brief discussion of the circuitry
that we devised to address these issues.
An extended discussion of the electronics appears in reference \cite{gianluA}.

\subsection{Thermistor biasing}

\begin{figure}
 \begin{center}
 \includegraphics[width=0.95\textwidth]{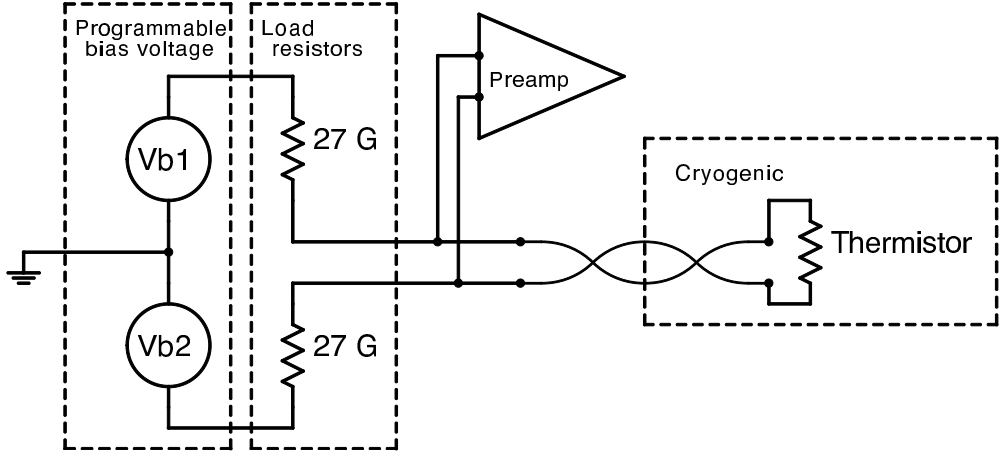}
 \end{center}
 \caption{Functional block diagram of the front-end electronics.
          Normally $R_L$ is 27 \gohm\ .
          For determination of a bolometer I-V characteristic 
          a command from the computer can reduce $R_L$ to 5.5 \gohm\ .}
\label{bias}
\end{figure}

We achieve thermistor biasing by the application of a programmable
voltage across a series combination of the thermistor and two 
load resistors. 
If electronic system is operated only at room temperature, 
the load resistors are also at room temperature ($R_L$ in Fig.~\ref{bias}).
The contribution to 1/f noise of large resistors under bias and the
minimization of this noise source is the subject of an independent
report \cite{gianluB}.
Under computer control we can reduce the biasing resistors from 
27 \gohm\ to 5.5 \gohm\, which facilitates a measurement of the 
bolometer's I-V characteristics at large bias.
The impedance at the input of the electronics is sufficiently high that
parasitic conductance in the circuit board is an issue that
demands great attention. 
Precautions that we took to suppress parasitic conductance were 
a) segregation of inverting and non-inverting inputs by circuit board layer, 
b) strategic placement of guard traces around the solder junctions of
high-impedance component leads,
c) maintenance of a minimum spacing of $5\sim mm$ between high-impedance
traces, and
d) the placement of slots at critical locations on the circuit board to
avoid conductance through minute amounts of soldering residue that
sometimes escapes the standard cleaning procedures.
Together these measures yield excellent results. 

If a cold buffer stage will be adopted, the pair of load resistors $R_L$ of 
Fig.~\ref{bias} will be moved close to the location of the buffer itself, at low temperature. 
Their value will be 27 \gohm\ . In this case the rest of the biasing set-up will remain unchanged.

The application of bias voltage to the load resistors is the
task of the programmable bias attenuator.
This circuit is a five-bit R-2R ladder network with optical isolation of
the digital input.
Its  outputs are the voltages $V_{b1}$ and $V_{b2}$ appearing in Fig.~\ref{bias}.
They take the value
 \begin{displaymath}
    V_{b1} = -V_{b2} = (n/64)(V_+-V_-)
 \end{displaymath}
where $V_+$ and $V_-$ are derived from an Agilent Technologies 6627A
system power supply common to all modules, $|V_+-V_-| $
may be set as high as 60 $V$ and $n$ can be set to any integer value between 1 and 64.
Interposed between the programmable bias attenuator and the load
resistors (but not shown in Fig.~\ref{bias}) is a bistable DPDT relay,
which provides for inverting the polarity of the bias voltage.
By reversing the bias voltage during in situ DC characterization, we can
account for the thermoelectric effect operating in the leads that connect
the front-end electronics to the thermistors.
Heavy filtering distributed through the biasing network reduces to a
negligible level whatever noise originates from the power system or from
the network itself.

\subsection{Preamplifier}

The design of our preamplifier was influenced by four paramount
considerations.
First, for the source impedance and signal characteristics that the
bolometer presents, a voltage amplifier is the conventional approach
\cite{mather_jones}.
Its performance is not superior to a charge or current configuration
\cite{gianluE}, but it allows its inputs to float.
Floating inputs make biasing compatible with DC coupling.
Second, the leads that connect the preamplifiers
with the bolometers operating near 10 $mK$ can extend between 1 $m$ and 5 $m$ in length.
The drawback of these long leads is that they act as antennas for
microphonic noise.
We address this issue by making the preamplifier input differential.
Compared with a single-sided input the amplifier noise power is double,
but for the more troublesome microphonic noise originating in the leads,
which is common mode, we achieve efficient suppression.
The differential configuration also serves to suppress interchannel
crosstalk.
Third, the signal bandwidth of our bolometers extends from 1 $Hz$ to about
10 $Hz$, and we ask that the amplifier noise, both series and parallel, be
minimized in this frequency interval.
Fourth, monitoring the baseline of each bolometer and correlating
baseline shift with shifts in overall system gain require low thermal 
drift in the preamp.

In addition to the above features the use of a cold buffer stage can increase further 
the rejection to microphonism by minimizing the high impedance section of 
detector's connecting leads.

\begin{figure}
 \begin{center}
 \includegraphics[width=0.95\textwidth]{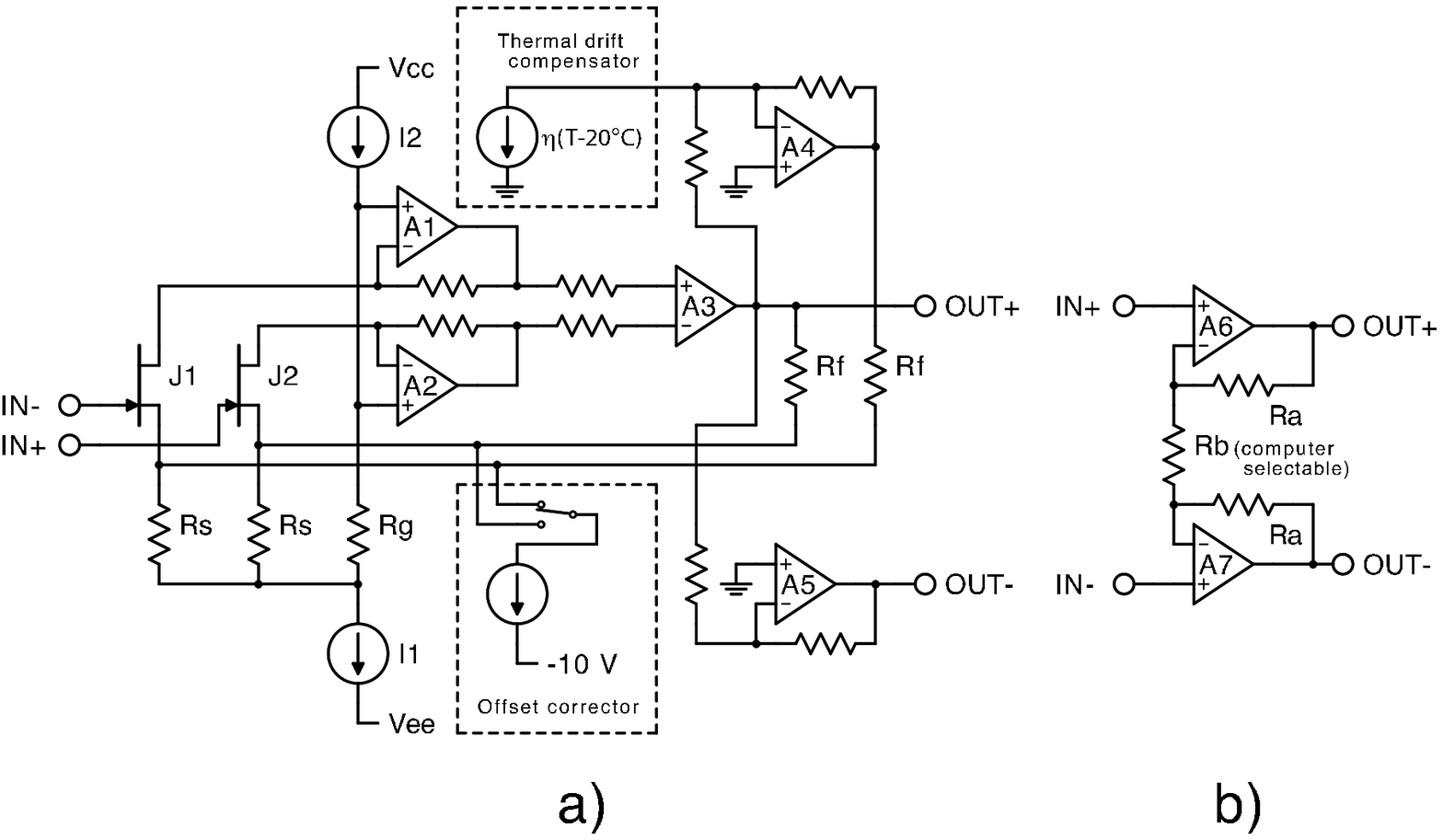}
 \end{center}
 \caption{Schematics of a) the preamplifier and b) the programmable gain amplifier. }
\label{preamp}
\end{figure}

To address these considerations the room temperature version of the front-end 
we adopted is the circuit shown in
Fig.~\ref{preamp}a.
At the core of the design is the JFET pair J1 and J2, which accepts the
differential input.
We selected the transistor type for these JFETs with a view to
minimizing the noise, especially at low frequency.
The capacitance of the gate  of the selected transistor is about 80 $pF$,
which optimizes the tradeoff between low series noise, low parallel
noise, and matching to the detector.
To minimize parallel noise we establish for each of the JFETs 
VDS $= I_2R_g-(I_1-I_2)\,R_s/2$ at 0.8 $V$,
the lowest value consistent with adequate gain.
The connection of the outputs of the operational amplifiers (op-amp) A3 and A4 
through the resistors
Rf to the sources of the JFETs closes the feedback loop  and establishes
the gain at (Rf+Rs)/Rs.
With Rf = 20 $k$\ohm\ and Rs = 91 \ohm\, the gain is 220 $V/V$.
Neither the input JFETs nor any of the op-amp circuits  are
individually compensated  for thermal drift.
We deal with this issue by trimming the circuit as a whole.
One component of the current source in the thermal drift compensation
circuit of Fig.~\ref{preamp} a is a diode that serves as a thermometer.
The magnitude of the coefficient $\eta$ is proportional to the value of a
single resistor, and placement of a shunt determines the sign of $\eta$.
We operate the preamp in a temperature controlled chamber and trim
$\eta$ to neutralize the uncompensated thermal drift.
In this way we were able to reduce an uncompensated drift as large as
60 $\mu V/^\circ$C to about 0.1 $\mu V/^\circ$C.

To minimize further the microphonic noise coming from the lenght of the leads, we have also realized, 
and are testing, a solution that uses an additional stage of amplification, operated cold, inserted 
near the 4.2 $K$ plate of the dilution refrigerator (and therefore much closer to the detectors). 
Each cold stage is a differential buffer 
composed by two Silicon JFETs that work in source follower configuration. This way the 
length of the leads that connect the detector to this additional buffer stage is reduced 
to about 1 $m$, while the remaining part of the wires toward the room temperature is driven 
by the low output impedance of the buffer. The buffer stage works at the optimum temperature 
for Silicon, around 110 $K$. Each cold buffer, together to the load resistors, is located on a 
boards that accommodates 6 channels. The boards are realized on a substrate made of 
PTFE 25\% reinforced fibreglass. This solution allows obtaining two goals. From one side the 
small thermal conductance of the PTFE minimize the thermal power to be injected for increasing 
the temperature of the JFETs to the optimum one, starting from the heat sink at 4.2 $K$. On the 
other side the radioactivity level of PTFE 25\% reinforced fibreglass is better by a factor of 
about 2 with respect to the pure fibreglass. In addition the printed circuit boards have a thickness of only 0.5 mm in order to reduce the amount 
of mass present. 
Thermally shielded metallic boxes have been realized that are each able to contain 2 of such 
boards. Inside the boxes the boards are suspended by means of nylon wires that have
a small thermal conductance to the heat sink, at the 4.2 $K$. 
This solution allows a good reduction of noise (see Fig.~\ref{noise}) however it's realization 
presents some technical difficulties.

The input offset of the preamplifier, for whatever configuration, is the sum of contributions from 
the asymmetry of the JFETs and the bias voltage of the NTD thermistor.
The JFET asymmetry may contribute up to $\pm\;20\;mV$ to the offset.
In normal operation thermistor biasing contributes no more than 20 $mV$,
but for obtaining the thermistor I-V characteristic the bias voltage
should be settable to both polarities and to a magnitude as high as
50$mV$.
We designed the offset corrector circuit to compensate input offsets
from -80 $mV$ to +80 $mV$ under control of the digital section.
In a naive design for this circuit the central element would be a 
conventional CMOS digital-to-analog converter (DAC).
Such a design, however, would generate excessive low-frequency noise.
Our design incorporates a DAC8043A, which is less noisy than ordinary
DACs because it omits the conventional output buffer.
To derive adequate current from this device we place it in the feedback
loop of a low-noise op-amp.
We achieve further noise suppression by using the DAC8043A to swing the
output through just 20 $mV$, i.e. 1/4 of the maximum output excursion.
A simple arrangement of two analog switches in series with low-frequency
noise-free resistors provides the necessary additional 60 $mV$ of output
swing.

\subsection{Programmable gain amplifier}

The outputs of the preamp drive the differential inputs of the
programmable gain amplifier, which we  show in Fig.~\ref{preamp}b.
The computer can set $R_b$ to any parallel combination of a set of five
resistors, 31 possibilities.
The resulting gain from preamp input to amplifier output takes values
from 220 to 5000 $V/V$.
The differential outputs travel on an analog bus to a filter circuit, which
is followed by analog to digital conversion.

The supply voltage for the very front-end so far described has been designed for the purpose for 
having very good stability, of a few $ppm/C$, and large rejection ratio \cite{gianluF}. 
These implemented features provide great stability during a long time measurement. 
One such supply voltage is dedicated to 12 boards, or 24 complete channels.

\subsection{Antialiasing filter and analog trigger}

In this experiment it is feasible to apply optimal filtering algorithms to
the detector signal in the off-line analysis, and signal shaping in hardware 
is thus inappropriate.
To ensure adequate frequency response we implement in hardware only an
antialiasing filter at the downstream end of the analog signal 
processing.
This low-pass filter is an active six-pole Bessel design, which yields a
roll-off of 120 $dB/decade$.
To allow for adapting the bandwidth to the individual bolometer modules
the cutoff frequency is settable from the computer to 8, 12, 16 or 20 $Hz$.

The trigger circuit input is AC coupled to the output of the
antialiasing filter.
Its gain is settable from the computer to 4.2, 12.5, 21 or
29 $V/V$.
The AC coupling produces a negative excursion of the trigger circuit
input that, if uncompensated, would result in a shift of the baseline.
A baseline restoring section \cite{gianluC} in the trigger circuit
suppresses this effect.
Pulses from a bolometer exhibit a trailing undershoot because of the
coupling of the inductive part of the bolometer's dynamic impedance to
the parasitic capacitance in shunt with it.
After differentiation by the AC coupling, the undershoot of large pulses
has the potential to retrigger.
An RC network \cite{gianluD} within the trigger circuit cancels the
effect of the undershoot and thus suppresses retriggering.
Its time constant is settable from the computer to 14, 27, 53, or
66 $s$.
The final element in the trigger circuit is a two-pole Bessel low-pass
filter.
The cutoff  of this filter tracks the cutoff of the antialiasing filter
but is higher in frequency by 4 $Hz$.
One trigger-Bessel module accommodates three channels.

\subsection{Data acquisition and triggering}

The analog trigger and antialising circuits are situated in a small
Faraday cage close to the acquisition and triggering system to which
they send their outputs.
The data acquisition system (DAS) is based on the VXI standard. 
Its main goal is the implementation of a zero dead-time system based on
the continuous analysis of a large circular memory buffer characterized 
by a very fast data transfer rate (DMA) with the ADC memories.
The DAS consists of an embedded controller PC (DASC), and few trigger and
ADC cards.
Signals from the front-end electronics are sent both to a series of 16~bit
ADC modules (where they are digitized) and, after proper differentiation,
to simple threshold trigger modules. 
The ADC sampling interval is set to values in the range 1-10~$ms$ depending
on the observed characteristics of the signals. Because of these large
sampling times, each ADC module is multiplexed and can serve a number of
acquisition channels. Sampled data are continuously transferred to a large circular
buffer of the DASC. When a trigger signal is also transferred to the DASC the
circular buffer is analyzed in order to recognize and extract the desired sampled
interval withouth interfering with the ADC activities. In normal conditions such a
logic does not require any DAQ dead-time. Synchronization between trigger
and ADC modules and the DASC
is, of course, of crucial importance.
The trigger discriminating thresholds are settable in increments of 5~$mV$.
This threshold adjustment in combination with the gain selection in the
trigger circuit provides channel-by-channel fine control of the trigger
level.
Given the variability of the energy-to-voltage conversion among
bolometers, the ability to tailor the trigger level to the bolometer
module is a valuable feature particularly for what concerns dark matter
searches for which the trigger level should be as low as possible.

\subsection{Noise considerations}

When the cold buffer stage is not present, the analog electronics chain just described 
produces noise at the level
of 50~$nV$ (FWHM) referred to the input over the 10~$Hz$ bandwidth of the
signal.
The 27~\gohm\ load resistors produce 80~$nV$ in this bandwidth.
The combined noise is then about 100~$nV$ (FWHM). 
When the cold buffer stage is present the series power noise is increased a factor of about 2, 
while parallel noise is decreased by a factor of at least 3, mainly by the use of the cold load 
resistors. This way the overall noise contribution becomes about 90~$nV$ (FWHM).
In both cases the noise is considerably below the 1.0~$\mu V$ (FWHM) 
level of noise that we have observed from a reference bolometer that we
consider to be an exemplar of good performance.
On the basis of this assessment we believe that the analog front end is
more than adequate for the needs of the experiment.

The two front-end solutions have the same level of noise, but the presence of the cold buffer stage, 
reducing the high impedance path, should further improve the common mode noise from the vibration of the leads. 
Unfortunately there is another source of microphonism, that is not a common mode, due to the mechanical 
friction between the detector and its mounting. If this source will result in noise much greater than that 
of the connecting leads, the use of a cold buffer stage will be not necessary. This fact will save a factor of two in the number of connecting leads (two for the readout and two for biasing for 
the case of the cold stage), and will save thermal power injection inside the fridge. The savings in 
parasitc capacitance that is obtained with shortening the high impedance path is not a concern with 
the foreseen very slow signals.

\subsection{Digital control}

A digital control system (DCS) enables the computer to manipulate all of
the significant parameters of the analog front end.
These parameters include the thermistor bias current, the input offset
correction, the gain, and the application of test signals to the inputs.
During data acquisition the DCS is generally idle.
In this state even the DCS clock shuts down thus eliminating the
possibility that DCS activity will contribute to noise on the analog
outputs.

\section{Off line analysis} \label{sec:analysis}

When the output voltage of one detector exceeds the trigger threshold,
the acquisition system saves a number of converted points of
the corresponding channel. Accordingly, for each
electronic-pulse, the wave form corresponding to a time window of
$\sim$ two seconds is sampled and acquired. The existence of a pre-trigger 
interval guarantees that a small fraction of the number of points
acquired are measurements of the DC level of the detector, or
it's temperature, just prior to the production of the pulse.

The fact that for each trigger an entire wave form is sampled
and recorded, implies that off-line analysis of a large body of
data will require a significant effort. Nevertheless, this effort
will be justified because of the useful information that can
be extracted from the signal waveform. The following are important
goals for the analysis:
\begin{enumerate}
\item{maximization of the signal to noise ratio for the best
estimate of the pulse amplitude. This is accomplished
by means of the optimum filter (OF) technique \cite{egatti};}
\item{ability to correct the effects of system instabilities that
change the response function of the detectors;}
\item{ablility to reject spurious triggered pulses by means of
pulse shape analysis;}
\item{capability of identifying and rejecting radioactive
background pulses by means of coincidence analysis.}
\end{enumerate}
The optimum filter (OF) technique is frequently used with bolometers
to evaluate the amplitude of a signal superimposed on stochastic
noise. This algorithm has proven to provide the best estimate of
the pulse amplitude under general conditions. Relative
to a simple maximum-minimum algorithm, this technique allows
the evaluation of the signal amplitude with much higher efficiency
resulting in an improvement in energy resolution by as much as a 
factor of two. To use the OF technique, the following information 
is necessary: the
response function i.e., the shape of the signal in a zero noise
condition, and the noise power spectrum. Once these are known,
the OF function is determined. It is then used as a software
filter for the acquired pulses. The role of the OF function
is to weight the frequency components of the signal to suppress
those frequencies that are highly influenced by noise. The
amplitude of the pulse is then evaluated using optimally
filtered pulses. Some pulse shape parameters can be deduced
from the comparison of the filtered pulses with the filtered
response function.

In processing the data off line, the following parameters are
evaluated and recorded to disk for each digitized pulse:
\begin{enumerate}
\item{the \emph{channel number} i.e., the ADC channel that exceeded
  the trigger threshold;}
\item{the \emph{absolute time} at which the pulse occurs with a precision of 0.1 $ms$;}
\item{the \emph{OF amplitude} i.e. that of the optimally filtered signals}
\item{the \emph{baseline}, obtained by averaging a proper number of
  points from the pre-trigger interval. It is a direct
  measurement of the DC level at the output of the detector
  just prior to the production of the pulse, and therefore
  it provides a measurement of the detector temperature at
  the creation of the signal;}
\item{the signal \emph{rise} and signal \emph{decay times};}
\item{the \emph{pulse shape parameters}, obtained by comparison between
  the pulse and the expected response function of the bolometer.
  The algorithms used to make this comparison, and to produce a
  numerical evaluation of it, are based on two main techniques:
  the comparison of the wave forms of the pulse and response function
  after OF, or Adaptive Filtering. A third powerful technique is based
  on artificial neural networks;}
\item{the \emph{pile-up fraction}, pile up is usually efficiently rejected
  using pulse shape analysis; however, this does not guarantee the
  identification of the rejected event as one from a pile-up. To
  improve this rejection, and to be able to evaluate the fraction
  of piled-up events, the Wiener-Filter algorithm is implemented [34].}
\end{enumerate}
The next step in the off line analysis is gain instability correction.
The OF amplitudes are corrected to reduce or cancel the effects of
system instabilities responsible for the variation of the ratio between
the energy, $E$, absorbed by the crystal and the amplitude $\Delta V $ of the
corresponding electrical pulse. In a very naive model (see Eq.\ref{eq:pulse}) 
of bolometers the following relation exists:
\begin{equation}
     \Delta V =  G_{e} \;A\; V_{bol}\; \frac{E}{C\;T_{bol}} = 
     G_{e} \;A\; V_{bias}\; \frac{R_{bol}}{R_{bol}+ R_{load}} \,\frac{1}{C\;T_{bol}}\; E
\label{eq:deltaV}
\end{equation}
Where $ G_{e} $ is the total gain of the electronic chain, A~=~thermistor sensitivity, 
$ V_{bias} $~=~total bias, 
$ V_{bol} =V_{bias} \, (R_{bol}/R_{bol}+ R_{load})$ , 
$ R_{load} $~=~load resistance, $ R_{bol} $~=~bolometer resistance, 
$ C_{bol} $~=~crystal heat capacity, $T_{bol}$~=~temperature of the crystal.

There are three instabilities that can modify the ratio $ \Delta V/E $: a
variation in the electronic gain ($ G_{e} $), a variation in the bias ($ V_{bias} $ ), 
and finally a variation in the temperature ($T_{bol}$)
of the crystal. 
The electronic system is designed to guarantee a stability of $ G_{e} $ and 
of $ V_{bias} $ within 0.1\% . It is however, much more difficult to maintain 
stability within 0.1\% of the detector temperature on long time scales. 
At a temperature of 10 $mK$
this would require maintaining the temperature of 1000 crystals to
an accuracy of 10 $\mu K$ for a period of several days.

To overcome this problem, a silicon resistor glued to each crystal
is used as a heater to produce a reference pulse in the detector. It
is connected to a high precision programmable pulser that produces a fast voltage 
pulse every few minutes dissipating the same amount of energy 
($E_{ref}$) into the crystal each time. These voltage pulses mimic pulses produced 
in the crystal by particle interactions and are used to measure the value of the 
ratio $ \Delta V/E $. Any variation of the amplitude of the reference pulse is due to 
variations of the $ \Delta V/E $ ratio.  The OF amplitude of the reference pulse is 
therefore used to measure, every few minutes, the actual value of $ \Delta V/E $ 
while the baseline of the reference pulse provide the contemporary measurement 
of the value of T. A fit is then used to obtain the values of $ \Delta V/E $ for 
each temperature value.
Therefore, in this step of the off line analysis, the OF amplitude of each
pulse is corrected according to the value assumed by the $ \Delta V/E $ for the 
detector temperature at which the pulse has been generated.
This technique has been proven to be efficient in the MI-DBD  experiment, where a typical 
temperature fluctuation over a day ranged from a few tenths to $ \sim 100 \; \mu K $.
After correction, these fluctuations were reduced to less than $ 1\; \mu K $ \cite{stabilizz}.

Pulse shape analysis is very useful in rejecting spurious signals produced
by microphonics and electronic noise. A confidence level is determined for
each pulse shape parameter and for the rise and decay time of each pulse.
Signals falling within these intervals are defined as "true" pulses, and
signals having one or more of their parameters outside of the relevant interval
are rejected as noise. The use of more than one pulse shape parameter results
in better reliability of the rejection technique.

The linearization of the detector response is critically important for
energy calibration. The final step in data processing is the conversion
of the OF amplitudes into energy values. The naive bolometer model
previously used assumes linearity; however, several parameters depend
on the crystal temperature, rendering the corresponding equation non-linear. Accordingly,
the relation between $ \Delta V $ and $E$ will be obtained periodically by the use of
radioactive calibration sources. The ratio $ \Delta V/E $ will be measured for several
gamma lines, and the data will be fit to the model previously described, but
taking into consideration the fact that the bolometer resistance and the crystal
heat capacity are temperature dependent. This will provide the calibration
function of $E$ as a function of $ \Delta V $, that will then be used to convert the OF
amplitudes into energy values.

Finally, the close packed array of the crystals will allow the rejection
of events that left energy in more than one single crystal. This will be
particularly useful in rejecting a very high energy gamma rays that enter
from outside of the array. The small angle Compton scattering in a single
crystal can mimic a double beta decay event, a dark matter scattering,
or a solar axion. The probability that this photon would escape the
rest of the array without a second interaction is small. In addition,
background events from radioactivity within the structure of the array
will also have a significant probability of depositing energy in more
than one crystal. This will also be true for high and intermediate energy
neutrons. In the final stage of off-line analysis these coincidence events
can be identified from the data which will contain the detector number,
signal time, pulse energy, and pile-up parameter.

\section{Simulation and predicted performances} \label{sec:simulation}

The goal of CUORE is to achieve a background rate in the range 0.001 to 
0.01 $ counts/(keV\;kg\;y)$ at the $0\nu\beta\beta$ transition energy of $^{130}Te$ 
($2528\;keV$). A low counting rate near threshold (that will be of the order of $\sim5\;keV$) 
is also foreseen and will allow CUORE to produce results in the Dark Matter and Axions research fields.
In Sec.~\ref{sec:bkgsim} we present a very conservative evaluation of the background reachable 
with CUORE; this is mainly based on the status of the art of detector design and of radioactive 
contaminations. This rather pessimistic approach in background evaluation is the only one that 
presently guarantees a reliable prediction. However CUORE construction will require about five 
years, there will therefore time enought for an $R\& D$ dedicated to background reduction as it is 
forseen in most next generations experiments. In Sec.~\ref{sec:bkgRD} we discuss this possibility.

\subsection{Background simulations} \label{sec:bkgsim}
Radioactive contaminations of individual construction materials, as well as the laboratory 
environment, were measured and the impact on detector performance determined by Monte Carlo 
computations. 

\begin{table} 
\caption{Bulk contamination levels in picograms per gram.} \label{contaminazioni}
\begin{center}
\begin{tabular}{|c|c|c|c|c|c|}
  \hline
  Contaminant & $^{232}Th$ & $^{238}U$ & $^{40}K$ & $^{210}Pb$ & $^{60}Co$ \\
  \hline
  $TeO_2$ & 0.5 & 0.1 & 1 & 10 $\mu$Bq/kg & 0.2 $\mu$Bq/kg \\
  copper & 4 & 2 & 1 & 0 &  10 $\mu$Bq/kg\\
  Roman lead & 2 & 1 & 1 & 4 mBq/kg & 0 \\
  16 Bq/kg lead & 2 & 1 & 1 & 16 Bq/kg & 0 \\
  \hline
\end{tabular}
\end{center}
\end{table}
The code is based on the GEANT-4 package; it models the shields, the cryostat, the detector structure 
and the detector array. It includes the propagation of photons, electrons, alpha particles and heavy 
ions (nuclear recoils from alpha emission) as well as neutrons and muons.
For radioactive chains or radioactive isotopes alpha, beta and gamma/X rays emmissions are 
considered according to their branching ratios. The time structure of the decay chains is taken into 
account and the transport of nuclear recoils from 
alpha emissions is included.

The background sources considered are:
\begin{enumerate}
\item{bulk and surface contamination of the construction materials in $ ^{238}U $, $ ^{232}Th $ chains and
$ ^{40}K $ and $ ^{210}Pb $ isotopes;}
\item{bulk contamination of construction materials due to cosmogenic activation;}
\item{neutron and muon flux in the Gran Sasso Laboratory;}
\item{gamma ray flux from natural radioactivity in the Gran Sasso Laboratory;}
\item{background from the $2\nu\beta\beta$ decay.}
\end{enumerate}

 For bulk contaminations the main contributions to background come from the heavy 
structures near the detectors (the copper mounting structure of the array, 
the Roman lead box and the two lead disks on the top of the array) and from the detectors 
themselves (the $TeO_2$ crystals). The radioactivity levels used in the computations 
for these materials are given in Table~\ref{contaminazioni}.
 The $^{232}Th$, $^{238}U$, $^{40}K$ and $^{210}Pb$ contamination levels of $TeO_2$, copper and lead as well
 as the $^{60}Co$ cosmogenic contamination of copper are deduced from the 
90\% C.L. upper limits obtained for the contaminations measured in 
MI-DBD experiment and with low activity $Ge$ spectrometry \cite{heusser}. 
In both cases no evidence of a bulk contamination is obtained with the achievable sensitivity 
and only upper limits could be obtained. 
The $^{60}Co$ cosmogenic contamination of $TeO_2$ is on the other hand evaluated according to the time
required to produce $TeO_2$ powder from the ore, grow the crystals and store them underground.
 The results of the Monte Carlo simulations using these contamination levels are given in 
Table~\ref{bulk} for the $0\nu\beta\beta$ decay and the low 
energy (10-50~$keV$) regions. The threshold assumed is 10 $keV$ and only values after 
reduction by anti-coincidence between detectors are there indicated.
 These values have to be considered as upper limits on the possible contribution of bulk 
contaminations to the CUORE background; they prove that with the already available materials  
background lower than $\sim 4\times10^{-3}\;counts/(keV\;kg\;y)$ in the $0\nu\beta\beta$ decay region 
and $\sim 3\times10^{-2}\;counts/(keV\;kg\;d)$ in the low energy (10-50~$keV$) region are ensured.
 
\begin{table}
\caption{Computed background in the $0\nu\beta\beta$ decay and in the low energy
regions for bulk contaminations in the different elements, the $Cu$ structure accounts for the detector mounting
structure and the 50~$mK$ shield.} \label{bulk}
\begin{center} 
\begin{tabular}{|c|c|c|c|c|}
  \hline Simulated & $TeO_2$ & $Cu$ & $Pb$ & TOTAL \\
  element & crystals & structure & shields &  \\
  \hline $0\nu\beta\beta$ decay region & & & & \\
  $counts/(keV\;kg\;y)$  & 1.6 $\times 10^{-3}$ & 1.5 $\times 10^{-3}$ 
  & 7.0 $\times 10^{-4}$ & 3.8 $\times 10^{-3}$\\
  \hline dark matter region & & & &  \\
  $counts/(keV\;kg\;d)$ & 2.3 $\times 10^{-2}$ & 9.6 $\times 10^{-4}$ 
  & 5.0 $\times 10^{-5}$  & 2.4 $\times 10^{-2}$\\
  \hline
\end{tabular}
\end{center}
\end{table}

Surface contaminations contribute to background only when they are localized on the crystals or on the
copper mounting structure directely faced to them. Background contributions in the two regions of 
interest are summarized in 
Table~\ref{surf}. There U and Th contaminations with an exponential density profile and
a $\sim$~1~$\mu m$ depth (in agreement with the contaminations observed in the MI-DBD experiment) are
considered, while the contamination of this surface layer is assumed to be: for crystals $\sim$~100 
times lower and for copper $\sim$~50 times lower than the corresponding contamination in MI-DBD 
experiment.
As discussed in Sec.~\ref{sec:radioac} the MI-DBD crystals were contaminated during the polishing 
procedure due to the use of highly contaminated powders. Polishing powders with a radioactive 
content 1000 times lower are however commercially available \cite{sumitomo} and have already been used for 
Cuoricino, an improvement of the surface contamination of a factor 100 is therefore fully
justified. A similar situation is that of copper; in the MI-DBD experiment the copper surfaces were 
treated with an etching procedure \cite{palmieri} shown to reduce impurities on surfaces before the 
sputtering process. 
This procedure allowed to obtain a good improvement of the surface quality of copper however it is 
not optimized from the point of view of background. The use of low contaminated liquids (water and 
inorganic acids are available with U and Th contaminations lower than 0.1~$pg/g$) in a low background 
environment will allow an improvement of one to two orders of magnitude in the surface contamination of 
copper.

\begin{table}[b]
\caption{Computed background in the $0\nu\beta\beta$ decay and in the low energy
regions for surface contaminations.}  \label{surf}
\begin{center} 
\begin{tabular}{|c|c|c|c|}
  \hline Simulated element& $TeO_2$ crystals & $Cu$ mounting structure & TOTAL \\
  \hline $0\nu\beta\beta$ decay region & & & \\ 
  $counts/(keV\;kg\;y)$  & 8.5 $\times 10^{-4}$ & 2.0 $\times 10^{-3}$ & 2.8 $\times 10^{-3}$  \\
  \hline dark matter region  & & & \\
   $counts/(keV\;kg\;d)$ & 6.8 $\times 10^{-4}$ & 5.3 $\times 10^{-4}$ & 1.2 $\times 10^{-3}$\\
  \hline
\end{tabular}
\end{center}
\end{table}

 Concerning the other background sources i.e. cosmogenic activation as well as 
muons, neutrons and gamma rays from the Laboratory environment these should produce minor
contributions to the $0\nu\beta\beta$ background thanks to the underground storage of construction 
materials and the optimization of the lead and neutron shields. In particular the only cosmogenically
produced long lived isotope contributing to $0\nu\beta\beta$ background is $^{60}Co$ whose presence has been
already included as a bulk contamination of both $TeO_2$ and copper(see Table~\ref{contaminazioni} and 
Table~\ref{bulk}). The situation is a little bit different for what concerns the low energy region. 
A preliminar evaluation of the non-intrinsic background rates near threshold results in a 
counting rate similar to that due to bulk contaminations. A complete and detaild study of the 
non-intrinsic background rates is undeway.

Using the present upper limits for the $^{130}Te$ $2\nu$~half-life,
the unavoidable background produced by the $2\nu\beta\beta$ decay is lower than 
10$^{-4}$~$counts/(keV\;kg\;d)$
(using the predicted energy resolution).

In summary, in a very conservative approach where the CUORE-tower mechanical structure is assumed identical 
to the
structure used in CUORICINO, with the presently achieved quality of low contamination materials and considering
the worst possible condition for bulk contaminations (i.e. all 
the contamination equal to the present 90\%/,C.L. measured upper limits (Table~\ref{contaminazioni}), the
CUORE background will be $\sim 0.007\,counts/(keV\;kg\;y)$ at the $0\nu\beta\beta$ transition 
and $\sim 0.05\,counts/(keV\;kg\;d)$ near threshold. 

\subsection{Background $R\,\& D$ for CUORE} \label{sec:bkgRD}

CUORE construction will require about five years; during that period we plan to spend much effort in trying 
to further reduce the achievable background. There are indeed several points on which it is possible to work.
Concerning the crystals, their 1.8 $\times 10^{-3}\,counts/(keV\;kg\;y)$ contribution to the $0\nu\beta\beta$ region 
is  dominated by the $^{60}Co$ cosmogenic contamination. This could be lowered by one order of magnitude by producing 
the crystals in a shallow depth laboratory. For U and Th impurities we have only upper limits of order 
$10^{-13}\,g/g$. However we know that before crystallisation the $TeO_2$ powder is much more contaminated, 
therefore the selection of a lower contamination powder toghether with the efficient reduction of 
impurities by the crystallization procedure should guarantee a contamination of $~10^{-14}\,g/g$.
The situation for copper is a little bit different as not only it is possible to work on the selection of 
low contamination copper but it is also possible to reduce the amount of copper used for the mounting
structure of the detectors. 
This is particularly true in the case of the two copper frames (see Fig~\ref{piano}) that hold the crystals in the 
four-detector module, presently in CUORICINO (and also in our simulation) they are over sized from both the 
thermal and the mechanical  points of view. A reduction of at least a factor 2 of their mass and surface is 
possible with a consequent reduction of the bulk and surface contribution to background.
The background coming from surface contaminations is dominated, in the $0\nu\beta\beta$ region, 
by the copper contribution and mainly by the 4-crystal copper frames. This contribution could be reduced 
not only by the reduction of the frame mass but also by covering the copper sufaces with cleaner materials.
If succesfull this $R\,\& D$ would then result in a reduction of background to the level of 0.001 
$counts/(keV\;kg\;y)$ at the $0\nu\beta\beta$ transition energy and to 0.01 
$counts/(keV\;kg\;d)$ at threshold.

\subsection{CUORE $0\nu\beta\beta$ sensitivity}

The connection between the effective electron neutrino mass and the nuclear structure 
factor $ F_{N} $ is given in equation (\ref{eq:mnu}). The values of $ F_{N} $ calculated 
for $ ^{76}Ge$ and $ ^{130}Te $ are given in Table \ref{FN} for 11 different 
nuclear structure models.

\begin{table}
\caption{Computed values of 
$ F_{N} \equiv G^{o\nu}\,|M^{0\nu}_{f} - (g_{A}/g_{V})^{2} M^{0\nu}_{GT}|^{2}$ for $^{76}Ge$ and
$^{130}Te$
and the $<m_{\nu}>\; ^{130}Te$ sensitivity for $\tau_{1/2}=2\times 10^{26}_{y}$} 
\begin{center}
 \begin{tabular}{|c|c|c|c|c|}
 \hline
  $F_{N}(^{76}Ge)$ & $\frac{F_{N}(^{130}Te)}{F_{N}(^{76}Ge)}$ & $F_{N}(^{130}Te)$ & 
  $<m_{\nu}>$ &   \\
  \hline
  $1.12\times10^{-13}$ & 4.76 & $ 5.33\times10^{-13}$ & 0.050 &\cite{staudt} \\
  $1.12\times10^{-13}$ & 4.32 & $ 4.84\times10^{-13}$ & 0.052 &\cite{tomoda} \\
  $1.87\times10^{-14}$ & 21.20 & $ 3.96\times10^{-13}$ & 0.057 &\cite{vogel} \\
  $1.54\times10^{-13}$ & 10.58 & $ 1.63\times10^{-12}$ & 0.028 &\cite{weak} \\
  $1.13\times10^{-13}$ & 9.74 & $ 1.1\times10^{-12}$ & 0.035 & \cite{engel} \\
  $1.21\times10^{-13}$ & 4.13 & $ 5.0\times10^{-13}$ & 0.051 & \cite{civita} \\
  $7.33\times10^{-14}$ & 4.09 & $ 3.0\times10^{-13}$ & 0.067 & \cite{pantisA} \\
  $1.42\times10^{-14}$ & 8.73 & $ 1.24\times10^{-13}$ & 0.103 & \cite{pantisB} \\
  $5.8\times10^{-13}$ & 5.48 & $ 3.18\times10^{-12}$ & 0.020 & \cite{staudt-kuo} \\
  $1.5\times10^{-14}$ & --- & --- &   & \cite{large}\\
  $9.5\times10^{-14}$& 3.79 & $ 3.6\times10^{-13}$ & 0.060 & \cite{opera} \\
  \hline
 \end{tabular}
\label{FN}
\end{center}
\end{table}

It is evident from the values in Table \ref{FN} that for a given half-life sensitivity, 
a $ ^{130}Te $ detector can be between a factor of $ \sim 2 $ and  $ \sim 4.6 $ as 
efficient in probing the Majorana mass parameter, $ < m _{\nu}> $ as a $^{76}Ge$ detector.
However, because of the ratio of the molecular weights, the ratio of the isotopic abundances 
in the detector materials, there are 5.41 times as many $ ^{76}Ge $  atoms in a kg of $Ge$ metal isotopically 
enriched to  86\%  in $ ^{76}Ge $, as there are $ ^{130}Te $ atoms in a kg of natural 
abundance $ TeO_{2}$. If we average the ratios of the values in Table \ref{FN}, and drop the two highest,
and two lowest values, we find $ < F_{N}( ^{130}Te)/F_{N}(^{76}Ge)>\simeq 6.2 $. Considering the 
uncertainty in this procedure, we estimate that a detector of $ TeO_{2} $ of natural abundance is 
equivalent in decay rate to one of  86\%  $\; ^{76}Ge $ of the same mass.

CUORE will have 206 kg of $ ^{130}Te $ which will contain $ 1.19 \times 10^{27} $ atoms of
$ ^{130}Te $. If we assume, from our discussion of backgrounds, a counting rate of 0.01~$counts/(keV\;kg\;y)$ 
and an energy resolution of $\sim 5\,keV$ there will be less than 60 background counts/y to a 95\%  kg 
in the 5~$keV$ window. The half-life sensitivity would be:
\begin{equation}\label{eq8}
    T_{1/2}^{0\nu}(^{130}Te)\geq (8.25\times 10^{26})\cdot \left( \frac{t}{2\sqrt{BG}}\right) y.
\end{equation}
The total number of background counts expected in the 5~$keV$ window in 10 years is $\sim$ 460. 
The $ 1\sigma $  statistical fluctuation, $ \sigma = \sqrt{n} $, is $ \sim $ 21 and the $ 2\sigma $ C.L. 
is 42. The half-life sensitivity is then $ \sim 2\times 10^{26}\,y $.
Using equation \ref{eq:mnu}, we quote the sensitivities of the Majorana mass parameter in Table~\ref{FN}
in the last column.

In the case a background of 0.001 $counts/(keV\;kg\;y)$ would be reached the 10 year sensitivity would 
be $ \sim 3.5\times 10^{26}\,y $ (at 95\% C.L.) and the corresponding sensitivities of the Majorana 
mass parameter will be  $\sim$ 1.8 times higher than those quoted in Table~\ref{FN}
(i.e. in the range $11-56\,meV$).

\subsection{CUORE dark matter and axion sensitivity}

\begin{figure}
\begin{center}
\includegraphics[width=0.95\textwidth]{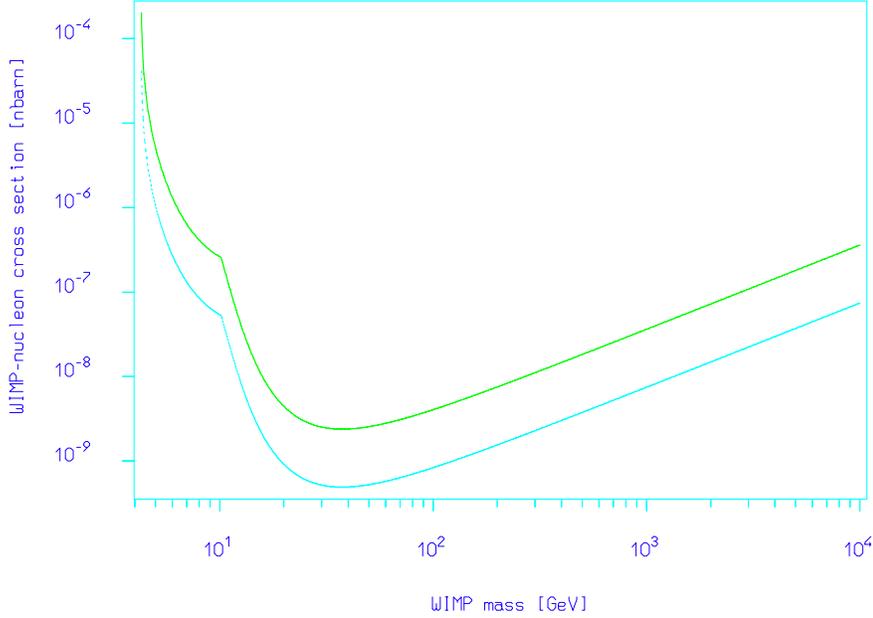}
 \caption{Exclusion plots for a background of 0.05 and 0.01 $counts/(keV\;kg\;d)$. }
\label{darkmatter}
\end{center}
\end{figure}

The sensitivity of CUORE in WIMP detection is shown in Fig~\ref{darkmatter} where the one year projected 
exclusion plots
for background counting rates of 0.05 and 0.01 $counts/(keV\;kg\;d)$) and a threhsold of 10 $keV$ 
are shown. Thanks to the large mass of the detector, CUORE will be also able to give results on the 
dark matter modulation signal as discussed in reference \cite{dm-morales}.
Concerning solar axion detection the sensitivity of CUORE will depend strongly on the threshold
reached. In the case of a 5~$keV$ threshold and a 2~$keV$ FWHM low energy resolution the 
$g_{\gamma\gamma}^{lim}$ sensitivity is of about $0.7\,10^{-9}\;GeV^{-1}$ \cite{creswick}.

\section{The enrichment option}\label{sec:enrich}
CUORE is the only future proposal for a $0\nu\beta\beta$ decay experiment that plans the use of
non-enriched material. Indeed as already the high isotopic abundace of $^{130}Te$ allows excellent 
high sensitivities even using the much cheeper natural abundance material. This obviously implies that the
construction of a larger array would be not only technically possible but also fiscally acceptable. On
the other hand the high natural abundance of $^{130}Te$ allows a much simpler and less expensive
enrichment procedure than that for other double beta decay candidates. Moreover $TeO_{2}$ crystals 
made with $^{130}Te$ isotopically enriched material have been already tested during the MI-DBD experiment with
results compatible with those obtained for the natural crystals both from the detector performances 
and detector background point of view.
An enriched CUORE with a background of 0.001 $counts/(keV\;kg\;y)$ would have a 10 years
half-life sensitivity of $ \sim 10^{27}\,y $ (at 95\% C.L.) implying a sensitivity on the Majorana mass
parameter in the range $9-46\,meV$.

\section{Conclusions}
 It is quite straightforward to demonstrate how the determination of the effective Majorana mass of the
electron neutrino constrains the lightest neutrino mass eigenvalue. This was done recently by Barger,
Glashow, Marfatia and Whistnant \cite{Barger}. However it is very convenient to use the approximation that
$sin\,\theta_{2} \simeq 0 \equiv s_{2}$ and $cos\,\theta_2 \simeq 1$. With the further assumption 
$\delta m_{\odot}^{2} \ll \delta m_{AT}^{2}$ in the normal hierarchy case we obtain:
\begin{equation}
|< m_\nu >| \,\le\, m_1\, \le\, \frac{|< m_\nu >|}{cos(2\theta_3)}
\end{equation}
where $cos(2\theta_3)\simeq 0.5$.
Similarly in the case of inverted hierarchy:
\begin{equation}
\sqrt{|< m_\nu >|^2-\delta m_{AT}^{2}}\, \leq\, m_1\,\leq\, 
\frac{\sqrt {|< m_\nu >|^2-\delta m_{AT}^{2}cos(2\theta_3)}}{cos(2\theta_3)} 
\end{equation}
A related quantity, $\Sigma \equiv m_1+m_2+m_3 $ has great importance in cosmology. It is related to the
quantity $\Omega_{\nu}$, the fraction of the critical density, $\rho_0$ (that would close the universe) that
is in the form of neutrinos, where:
\begin{equation}
\Sigma = (93.8\,eV)\; \Omega_{\nu}\;h^2
\end{equation}
and $h$ is the dimensionless Hubble constant. It was shown in reference \cite{Barger} that $\Sigma$ 
obeys the
following inequality with respect to $|< m_\nu >|$:
\begin{eqnarray}
\lefteqn {2\,|< m_\nu >|+\sqrt{|< m_\nu >|^2 \pm \delta m_{AT}^{2}} \leq \Sigma} \nonumber\\
& & \leq \frac{2\,|< m_\nu >|+\sqrt {|< m_\nu >|^2 \pm \delta m_{AT}^{2}\,cos(2\theta_3)}}{cos(2\theta_3)}
\end{eqnarray}
where the $+$ and $-$ signs refer to normal and inverted hierarchy respectively, and 
$cos(2\theta_3) \equiv 0.5$.
If this relation is written as two equalities, and each side is solved for $|< m_\nu >|$ in terms of
$\Sigma$, quadratic equations results with the constant terms $(\Sigma^2 \mp \delta m_{AT}^{2})$.
Values of $\Sigma^2$ of cosmological interest are much larger than $\delta m_{AT}^{2} \leq 0.005$ eV
(99.73\% C.L.). Accordingly the above inequality reduces to the trivial form:
\begin{equation}
|< m_\nu >|\, \leq \,\Sigma / 3\, \leq \, 2|< m_\nu >|
\end{equation}
This clearly demonstrates the direct connection between neutrinoless double beta decay and neutrino dark
matter. 
Cosmic microwave background and galaxy cluster surveys have set the bound $\Sigma \leq 1.8\,eV$ \cite{Wang}.
It is expected that the MAP satellite will produce data that will result in a sensitivity $\Sigma \simeq
0.5\,eV$. In the case of Majorana neutrinos, the CUORE experiment is projected to have a sensitivity od
$\Sigma \simeq (0.035-0.103)\,eV$ depending on the nuclear matrix element used to analyze the data.

One interesting scenario results in the case the MAP satellite, for example, observes a clear
singal well above their limiting sensitivity, and CUORE (as well as other next generation experiments)
have a negative results at sensitivities of $\Sigma$ far below that of the satellite data. This would be a
clear indication that the phenomenon causing the density fluctuations implied by the CMB data was not
caused by Majorana neutrino dark matter.
In addition a positive measurement of the mass $m_{\nu_e}$ by KATRIN $^3 He$ beta end-point spectrum
measurement would have to yield $m_{\nu_e} \geq 0.35\,eV$. The CUORE experiment will have a far greater
sensitivity and if it would find a negative result, the mystery of Dirac or Majorana character of neutrinos will
be solved.

It should be abundantly clear from the forgoing discussion that the most important remaining issue in
neutrino physics is that of the determination of the neutrino mass scale. Neutrino oscillation experiments
have clearly demonstrated that:
\begin{enumerate}
\item{neutrinos have mass and that their mass eigenstates mix}
\item{the mixing matrix elements have values that imply that next generation $0\nu\beta\beta$ decay 
experiments should be able to determine the scale of neutrino mass in the case of Majorana neutrinos.}
\end{enumerate}
The main advantage of CUORE is that it will utilize $TeO_2$ crystals of natural abundance Te, while this
technology has yielded energy resolutions approaching those of intrinsic Ge detectors. Energy resolution is
crucially important for an actual observation of $0\nu\beta\beta$ and a measurement of $|< m_\nu >|$.


\begin{thebibliography}{99}
\bibitem{furry}W. H. Furry, Phys. Rev. {\bf 56}, 1184 (1939).
\bibitem{primakoff}H. Primakoff and S. P. Rosen, Rep. Prog. Phys. {\bf 22}, 121 (1959); Phys. Rev. {\bf 184}, 1925 (1969); 
	G. F. Dell'Antonio and E. Fiorini, Suppl. Nuovo Cimento {\bf 17}, 132 (1960).
\bibitem{haxton}W. C. Haxton and G. J. Stevenson Jr., Prog. Part. Nucl. Phys. {\bf 12}, 409 (1984);
	 F. T. Avignone III and R. L. Brodzinski, Prog. Part. Nucl. Phys. {\bf 21}, 99 (1988); 
	 M. Moe and P. Vogel, Ann. Rev. Nucl. Part. Sci {\bf 44}, 247 (1994).
\bibitem{morales}A. Morales, Nucl. Phys. B (Proc. Suppl.) {\bf77}, 335 (1999);
	 H. Ejiri, Phys. Rept. C (2000), Int. J. Mod. Phys. E6, 1 (1997);
	 V. Tretyak and Y. Zdesenko, At. Data, Nucl. Data Tables {\bf 80}, 93 (2002).
\bibitem{elliottvogel} S. Elliott and P. Vogel, submitted to Ann. Rev. Nucl. Part. Sci., arXiv: hep-ph/0202264.
\bibitem{fiorini}E. Fiorini {\it et al.}, Phys. Lett. {\bf 25 B}, 602 (1967); Lett. Nuovo Cimento {\bf 3}, 149 (1970).
\bibitem{vasenko}A. A. Vasenko {\it et al.}, Mod. Phys. Lett. {\bf A 5}, 1299 (1990); 
  	F. T. Avignone {\it et al.}, Phys. Lett. {\bf B 256}, 559 (1991).
\bibitem{bandis}L. Baudis {\it et al.}, Phys. Rev. Lett. {\bf 83}, 41 (1999).
\bibitem{aalseth}C. E. Aalseth {\it et al.}, Phys. Rev. {\bf C 59}, 2108 (1999); 
	Yad. Fiz {\bf 63}, 1299, 1341 (2000) [ Phys. At. Nucl. {\bf 63}, 1225, 1268 (2000)];
        D. Gonzales {\it et al.}, Nucl. Phys. B (Proc. suppl.) {\bf 87}, 278 (2000).
\bibitem{ceaalseth}C. E. Aalseth {\it et al.}, Phys. Rev. {\bf D65}, 092007 (2002).
\bibitem{fukuda}Y. Fukuda {\it et al.}, Phys. Rev. Lett. {\bf 82}, 1810 (1999); 
		{\bf 82}, 2430 (1999); 82, 2644 (1999).

\bibitem{hampel}W. Hampel {\it et al.}, Phys. Lett. {\bf B 388}, 384 (1996), The GALLEX Collaboration;
		G. N. Abdurashitov {\it et al.}, The SAGE Collaboration, Phys. Rev. Lett. {\bf 77}, 4708 (1996);
		B. T. Cleveland {\it et al.}, Astrophysics J. {\bf 496}, 505 (1998);
		R. Davis, Prog. Part. Nucl. Phys. {\bf 32}, 13 (1994); 
		Y.Fukuda {\it et al.}, Phys. Rev. Lett. {\bf 86}, 5651 (2001); {\bf 86}, 5656 (2001).


\bibitem{ahmad}Q. R. Ahmad {\it et al.}, Phys. Rev. Lett. {\bf 87}, (2001) 071301; S. Fakuda {\it et al.}, arXiv:hep-ex/0205075; Phys. Lett.  {\bf B539}, (2002) 179.
\bibitem{pascoli}S. Pascoli, S. T. Petcov, and L. Wolfenstein, Phys.Lett. {\bf B524}, (2002) 319-331.
\bibitem{farzan}F. Feruglio, A. Strumia and F. Vissani, hep-ph/0201291; Nucl.Phys. {\bf B637}, 345 (2002).
\bibitem{bilenky}S. M. Bilenky, S. Pascoli and S. T. Petcov, arXiv: hep-ph/0105144 v1 15 May 2001; Phys. Rev. {\bf D64}, 053010 (2001).
\bibitem{mbilenky}S. Pascoli and S. T. Petcov, arXiv: hep-ph/0205022.
\bibitem{klapdor}H. V. Klapdor-Kleingrothaus, H. Päs, A. Yu. Smirnov, Phys. Rev. {\bf D 63}, 073005 (2001).
\bibitem{sbilenky}S. M. Bilenky, C. Guinti, W. Grimus, B. Kayser, and S. T. Petcov, Phys. Lett. {\bf B 465}, 193 (1999); Y. Farzan, O. L. G. Peres and Y. Smirnov, Nucl. Phys. {\bf B612}, 59 (2001).


\bibitem{mibetafinale} A. Alessandrello {\it et al.}, Phys. Lett. {\bf B 486}, 13 (2000);
	 C. Arnaboldi {\it et al.} "A calorimetric search on double beta decay of $^{130}Te$", submitted to Phys. Lett. B.

\bibitem{twerenbold}D. Twerenbold, Rep. Prog. Phys., {\bf 59}, 349 (1996);
 	 N. Booth, B. Cabrera, and E. Fiorini, Ann. Rev. Nucl. Sci., {\bf 46}, 471 (1996).
\bibitem{haller}E. E. Haller {\it et al.}, Proc. Fourth Int. Conf. on Neutron Transmutation Doping of Semiconductor Materials, Nat.
	Bureau of Standards, June 1,2 1982, Gaithersburg MD, R. D. Larrabee ed., (Plenum Press 1984) p 21.
\bibitem{norman}E. Norman and R. J. McDonald - Presentation made to the CUORE collaboration, 
		Milan Italy, 11 Decemter 2000 (Private communication).
\bibitem{dubna} A. Alessandrello {\it et al.}, Nucl. Phys. Russ. Academy of Sci, in press., arXiv: hep-ex/0201038

\bibitem{cuoricino} A. Alessandrello {\it et al.}, Nucl. Phys. B (Proc. Suppl) {\bf 87}, 78 (2000).

\bibitem{creswick}R. J. Creswick {\it et al.}, Phys. Lett. {\bf B 427}, 235 (1998);
	 F. T. Avignone III {\it et al.}, Phys. Rev. Lett. 5068 (1998); 
	 S. Cebrian {\it et al.}, Astropart. Physics {\bf 10}, 397 (1999).
\bibitem{bariucci}M. Barucci {\it et al.}, Journal of Low Temperature Physics (in press).
\bibitem{white}G. K. White, S. J. Collocott, and J. G. Collins, J. Phys. of Condensed Matter {\bf 2}, 7715 (1990).

\bibitem{giuliani}A. Giuliani and S. Sanguinetti, Mat. Sci. and Eng. {\bf R 11}, 52 (1993).
\bibitem{mott}N. F. Mott. Phil. Mag. {\bf 19}, 835 (1969).
\bibitem{shklovskii}B. I. Shklovskii and A. L. Efros, Sov. Phys. - JETP {\bf 33}, 468 (1971).
\bibitem{aalessandrello}A. Alessandrello {\it et al.}, J. Phys. D: Appl. Phys. {\bf 32}, 3099 (1999).
\bibitem{pobell}F. Pobell, "Matter and Methods at Low Temperatures", Springer - Verlag, Berlin, 1992.
\bibitem{rello}A. Alessandrello {\it et al.}, Proc. 7 Int. Workshop on Low Temperature Detectors (LTD7),Munich, Germany 27 July - 2 August 1997, ed. S. Cooper, Springei Veilag (1997) p 249.
\bibitem{drllo}A. Alessandrello {\it et al.}, Nucl. Insti. and Meth. {\bf B 142}, 163 (1998).

\bibitem{gianluA} C.~Arnaboldi {\it et al.}, "The Programmable Front-End System For Cuoricino, 
an Array of Large-Mass Bolometers", to be published on IEEE Transactions on Nuclear Science October 2002

\bibitem{gianluB} C.~Arnaboldi {\it et al.}, IEEE Transactions on Nuclear Science, to be published October 2002
"Low Frequency Noise Characterization of Very Large Value Resistors".

\bibitem{mather_jones} J.~C.~Mather, Applied Optics {\bf21} 1125-1129 (1982); R.C.~Jones, J. Opt. Soc. Am. {\bf43} 1-14 (1953)

\bibitem{gianluE} A.~Fascilla and G.~Pessina, Nucl. Instrum. Meth. A {\bf469} 116-126 (2001)

\bibitem{gianluF}G.~Pessina, The Review of Scientific Instruments, {\bf70} p.3473-3478, 1999

\bibitem{gianluC}C.~Liguori and G.~Pessina, Nucl. Instrum. Meth. A {\bf437} 557-559 (1999)

\bibitem{gianluD}C.~Arnaboldi and G.~Pessina, submitted to Nucl. Instrum. Meth. A, "A Very Simple Baseline Restorer for Nuclear Applications"

\bibitem{egatti}E.~Gatti and P.~F.~Manfredi, Rivista Nuovo Cimento, {\bf 9,1} (1986). 
\bibitem{stabilizz}A.~Alessandrello {\it et al.}, NIM A {\bf412} (1998) 454.
\bibitem{heusser}G.~Heusser, private communication.
\bibitem{sumitomo}Admatechs Co. Ltd., Japan.
\bibitem{palmieri}V.~Palmieri {\it et al.} "Electropolishing of seamless 1.5 GHZ OFHC copper cavities", 
	Proceed. of the $7^{th}$ workshop on RF Superconductivity, Gif Sur Yvette, France 1995, B. Bonin ed., DIST CEA/SACLAY 96 080/1,{\bf 2}, 605.
	 "Besides the standard Niobium bath chemical polishing";
	 Proceedings of the $10^{th}$ workshop on RF Superconductivity, Sept. 6-11  2001, Tsukuba, Japan, (2002) on press

\bibitem{staudt}QRPA, A. Staudt {\it et al.}, Europhys. Lett. {\bf 13} (1990) 31.
\bibitem{tomoda}QRPA, T. Tomoda, Rep. Prog. Phys. {\bf 54} (1991) 53.
\bibitem{vogel}QRPA, P. Vogel {\it et al.}, Phs. Rev. Lett. {\bf 57} (1986) 3148; Phys. Rev. {\bf C 37} (1988) 73.
\bibitem{weak}Weak Coupling S. M., W. C. Haxton {\it et al.}, Prog. Part. Nucl. Phys. {\bf 12} (1984) 409, and Nucl. Phys. {\bf B 31} (Proc. Suppl.) (1993) 82, Phys. Rev. {\bf D 26} (1982) 1085.
\bibitem{engel}Generalized seniority, J. Engel and P. Vogel, Phys. Rev. Lett. {\bf B 225} (1989)5.
\bibitem{civita}QRPA, O. Civitaresa, A. Faessler, T. Tomoda, Phys. Lett. {\bf B 194} (1987) 11; T. Tomoda, A. Faessler, Phys. Lett. {\bf B 199}  (1987) 473; J. Suhonen and O. Civitarese, Phys. Rev. {\bf C 49} (1994) 3055.
\bibitem{pantisA}QRPA without pn pairing, A. Pantis {\it et al.}, Phys. Rev. {\bf C 53} (1996) 695.
\bibitem{pantisB}QRPA with pn pairing, A. Pantis {\it et al.}, Phys. Rev. {\bf C 53} (1996) 695.
\bibitem{staudt-kuo}A. Staudt, Kuo and H. Klapdor, Phys. Rev. {\bf C 46} (1992) 871.
\bibitem{large}E. Caurier {\it et al.}, Phs. Rev. Lett. {\bf 77} (1996) 54, Retamosa {\it et al.}, Phys. Rev. {\bf 51} (1995) 371, A. Poves {\it et al.}, Phs. Lett. {\bf B 361} (1995) 1.
\bibitem{opera}Operator Expansion Method, J. G. Hirsch {\it et al.}, Nucl. Phys. {\bf A 589} (1995) 445, C. R. Ching {\it et al.}, Phys. Lett. {\bf B 272} (1991) 169, Phys. Lett. {\bf B 276} (1992) 274.
\bibitem{Barger}V. Barger, S. L. Glashow, D. Marfatia and K. Whisnant, Phys. Lett {\bf B 532} (2002) 15.
\bibitem{Wang}O.~Ergaroy {\it et al.}, Phys.Rev.Lett. 89 (2002) 061301, arXiv: astro-ph/0204152.
\bibitem{dm-morales}S.~Cebrian {\it et al.}, Astropart. Phys. {\bf9} 339 (2001), arXiv: hep-ph/9912394.

\end{thebibliography}
\end{document}